\documentclass[a4paper,11pt]{article}

\usepackage[toc,page]{appendix}
\usepackage{graphicx}
\usepackage{colortbl}
\usepackage{amsmath}
\usepackage{subfigure}
\usepackage{mathtools}
\usepackage[utf8]{inputenc}
\usepackage{float}
\usepackage[normalem]{ulem}
\usepackage{epstopdf}
\usepackage{amsfonts,amsmath,amssymb}
\usepackage{hyperref}
\usepackage{cite}

\usepackage{epsfig,multicol,bbm}
\usepackage{color}
\usepackage[dvipsnames]{xcolor}

\definecolor{darkblue}{rgb}{0.0, 0.0, 0.55}
\definecolor{grey}{rgb}{0.57, 0.64, 0.69}
\definecolor{lightbrown}{rgb}{0.71, 0.4, 0.11}
\newcommand{\tcb}{\textcolor{blue}}

\newcommand{\be}{\begin{equation}}
\newcommand{\ee}{\end{equation}}

\newcommand{\rbm}[1]{{\color{red}\bf [Robb: #1]}}

\newcommand{\fixme}[1]{{\color{red}{\bf{#1}}}}

\newcommand\fverb{\setbox\pippobox=\hbox\bgroup\verb}
\newcommand\fverbit{\egroup\item[\fbox{\unhbox\pippobox}]}

\newbox\pippobox

\begin{document}
\title{\bf More on Analytically Approximate Solution to Quadratic Gravity}
\author{Seyed Naseh Sajadi\thanks{Electronic address: naseh.sajadi@gmail.com}\,,\,Supakchai Ponglertsakul\thanks{Electronic address: supakchai.p@gmail.com}
%\small 
\\
\small Strong Gravity Group, Department of Physics, Faculty of Science, Silpakorn University,\\ Nakhon Pathom 73000, Thailand\\
%\small Department of Physics and Astronomy, University of Waterloo, Waterloo, Ontario, Canada N2L 3G1\\
%\small Physics Department and Biruni Observatory, College of Sciences, Shiraz University,
%Shiraz 71454, Iran
} 
\maketitle
\begin{abstract}
In this work, we obtain the analytically approximation of static, spherically symmetric black
hole solutions to Einstein$-$Weyl squared gravity by using the continued fraction expansion method. The black hole solutions are found for various relations between near horizon parameters with positive Weyl's coupling constant $\alpha$. Black hole solutions associated with different near-horizon constants are compared with the numerical ones. We obtain four branches of the black hole solutions with positive Arnowitt-Deser-Misner (ADM) mass. A non-Schwarzschild solution appears when the integration constant reaches a certain value for arbitrary values of the coupling constant {$\alpha$}. In addition, we study the thermodynamic and dynamical stability of the black hole solutions by considering thermodynamic quantities and the quasinormal frequencies.
\end{abstract}

\section{Introduction}
%\fixme{I will use the bib file instead for references. I'll handle this.}

Black holes (BH) are fundamental objects that arise as vacuum solutions in general relativity (GR). A static and spherically symmetric vacuum solution in GR gives rise to a Schwarzschild solution characterized by a horizon radius (or by its mass). In theories beyond GR, it is possible to realize non-Schwarzschild BH solutions whose geometries are modified by the presence of additional degrees of freedom. In scalar-tensor or vector-tensor theories, some asymptotically flat BH solutions endow with scalar or vector hairs. With the advent
of gravitational {waves} astronomy, we can now probe the physics of strong gravity regimes and the possible deviations from GR. On strong gravitational backgrounds, it is expected that higher-order gravity models may modify the spacetime structure and its dynamics. A {natural} alternative candidate to Einstein's theory in four dimensions is quadratic theories of gravity which was proposed in \cite{PhysRevD.16.953}. 
The Einstein quadratic gravity (EQG) field equations beside the Schwarzschild-like solution admit non-Schwarzschild black holes which were recently discovered in \cite{Lu:2015cqa, PhysRevD.92.124019}. These new analytical and numerical black holes have been studied in \cite{Rezzolla:2014mua, Podolsky:2018pfe}. In \cite{Bonanno:2019rsq} a complete analysis of the link between the weak field expansion, the structure of the horizons, and the interior geometry for static space-times has been presented. In \cite{Sajadi:2020axg, Sajadi:2022ybs, Sajadi:2022tgi}, following the numerical work by $L\ddot{u}$ et. al \cite{Lu:2015cqa}, the new analytical {asymptotically} flat black hole solutions have been obtained using continued fraction expansion. This method is a highly accurate analytic method {which} has recently been applied with success in a variety of contexts \cite{Kokkotas:2017zwt, Kokkotas:2017ymc, Konoplya:2022iyn,Antoniou:2024jku}. These analytic{al} studies of the black holes also allowed us to study the thermodynamics and dynamical stability of the solutions. 

Moreover, it is also important to investigate the physical properties of the novel black hole spacetime further. One can consider the ability to withstand a small perturbation. It is crucial to determine whether the spacetime is robust against a small perturbation. If it is not, a small perturbation may develop into a non-negligible backreaction on the spacetime itself and abruptly change the nature of the spacetime geometry. The study of a small perturbation on black holes can be done by considering quasinormal modes (QNMs) \cite{Konoplya:2011qq, Kokkotas:1999bd}. The QNMs are oscillation modes that satisfy certain boundary conditions i.e., only ingoing modes at the black hole's event horizon and only outgoing modes at infinity. These boundary conditions lead to discrete complex frequencies. The real part is interpreted as oscillation frequency while the imaginary part determines the damping time of each mode. For a spherically symmetric background, the fundamental equation governing a small perturbation on black hole spacetime is usually reduced to a single second-order ordinary differential equation. The perturbation equation is similar to an equation describing a particle encountering a potential barrier. There are several methods for solving the perturbation equation both analytical and numerical methods e.g., direct integration \cite{Promsiri:2023yda}, continued fraction method \cite{Leaver:1985ax}, asymptotic iteration method \cite{Cho:2009cj,Burikham:2017gdm,Ponglertsakul:2018smo,Ponglertsakul:2020ufm}, pseudospectral method \cite{Jansen:2017oag}, Wentzel–Kramers–Brillouin (WKB) method \cite{Schutz:1985km, Iyer:1986np, Konoplya:2003ii,Matyjasek:2017psv, Matyjasek:2019eeu, Konoplya:2019hlu,Tangphati:2023xnw,Gogoi:2024vcx,Bhar:2024ehw} etc. Among various methods, the Ferrari and Mashoon method \cite{vf} stands out as it allows one to obtain an exact solution in a simple form. The exact solution entirely depends on the height and curvature of the effective potential. The Ferrari and Mashoon methods have been utilized to obtain quasinormal frequencies in several spacetime backgrounds. For instance, QNMs of near-extremal Schwarzshcild de-Sitter black holes are obtained in \cite{Cardoso:2003sw}. In dRGT massive gravity, quasinormal frequencies are investigated using the Ferrari and Mashoon method for near extremal black hole and black string spacetimes \cite{Wuthicharn:2019olp} and later is extended to generalized spherically symmetric spacetime \cite{Burikham:2020dfi}. In \cite{Ponglertsakul:2020ufm}, the quasinormal spectrum of near extremal Myers-Perry de-Sitter black holes is computed. On the other hand, it is found that for near-extremal Kerr-Newmann-de-Sitter back holes, the Ferrari and Mashoon method is not applicable for the case of fermionic fields \cite{Churilova:2021nnc}.

The novelty of our results using this method is that \emph{for any values of coupling constant of theory, we obtain a non-Schwarzschild-like black hole solution. However, similar to the numerical work, we have four branches of the black hole solution with positive mass even for a linear relation between $f_{1}$ and $h_{1}$.} 

%The main objective of this paper is to extend the studies of \cite{Bonanno:2019rsq} and \cite{Sajadi:2020axg}. To be more precise, in the paper \cite{Bonanno:2019rsq} the authors show that the non-Schwarzschild-like black hole solution exists if there is a nonlinear relation between near horizon constants. However, we have demonstrated in \cite{Sajadi:2020axg}, that for the linear relation between the near horizon constants i.e., ($f_{1}=h_{1}$), the non-Schwarzschild-like black hole solution exists. 
%{In this paper, using the presented method we show that the numerical solutions can be obtained when considering the proper relationship between $f_1$ and $h_1$.}

The paper is organized as follows: in section \ref{sec2}, we present the basic formalism of the theory, continued fraction expansion method, and thermodynamics of the black holes. In subsection \ref{linearcase}, we have considered the case of $h_{1}=f_{1}$ to compare the analytical results with the numerical one with positive generic $\alpha$. The case $h_{1}= f_{1}^2$ is studied in subsection \ref{secc3}. In subsection \ref{sub22}, we study the dynamical stability of the {novel} solutions. In section \ref{conc}, we present our conclusions. 

\section{Basic equations}\label{sec2}
The general Lagrangian of Einstein's quadratic gravity can be written as 
\begin{equation}\label{eq1}
L=\gamma R-\alpha C_{a b c d}C^{a b c d}+\beta R^{2},
\end{equation}
where $ C_{a b c d} $ is the Weyl tensor 
whose square is given by $C_{abcd}C^{abcd}=R^2/3-2R_{ab}R^{ab}+R_{abcd}R^{abcd}$, $\beta $ and $ \gamma $ are coupling constants. { Since the Gauss-Bonnet curvature invariance is a topological term that does not affect the spacetime dynamics in four dimensions, the general gravitational action up to quadratic-order curvature terms is given by the action \eqref{eq1}.}
In theories without the Weyl curvature, i.e., $\alpha = 0$ in Eq. \eqref{eq1}, it has been shown that there exists a non-Schwarzschild solution \cite{Sajadi:2022tgi}. 

With the presence of the Weyl curvature term, the Schwarzschild BH is an exact solution on the static spherically symmetric background for 
arbitrary couplings $\alpha$. When the coupling $\alpha$ is of special order, it is known that the other asymptotically flat non-Schwarzschild branch appears beside the Schwarzschild branch\cite{Lu:2015cqa}. However, we have shown in \cite{Sajadi:2020axg}, that for an arbitrary coupling $\alpha$, when the constant of integration is of special value, the non-Schwarzschild branch emerges. {Moreover,} the negative $\alpha$ case has been studied in \cite{Bonanno:2019rsq}, showing there is no asymptotically flat solution for this case.

In the absence of a cosmological constant, the equation of motion does not get the contribution from the $\beta$ term. Henceforth, without loss of generality, we set $\beta=0$ and $\gamma=1$ \cite{Lu:2015cqa}. 
% Since the trace of the equations of motion without a cosmological constant vanishes, the term proportional to $\beta$ does not contribute to the solution. 
% Henceforth we set
% $\beta=0$, and  for simplicity we also set $ \gamma=1 $ \cite{Lu:2015cqa}.
The field equations are given by 
\begin{equation}
E_{a b}=R_{a b}-\dfrac{1}{2}g_{a b}R-4\alpha B_{a b}=0 ,\;\;\;\; B_{a b}=\left(\nabla^{m}\nabla^{n}+\dfrac{1}{2}R^{m n} \right)C_{a m b n}.
\end{equation}
We consider the following static, spherical symmetric metric
\begin{equation}\label{metform}
dS^{2}=-h(r)dt^{2}+\dfrac{dr^{2}}{f(r)}+r^{2}\left(d\theta^{2}+\dfrac{\sin^{2}(\sqrt{k}\theta)}{k}d\phi^{2}\right),
\end{equation}
{where $k$ can take the value of $\{-1,0,1\}$ which corresponds to the hyperbolic, spherical, and brane topology of the metric. With this metric, the trace of the field equation implies }
% By inserting the metric into the field equations we obtain the differential equations for $ f(r) $ and $h(r)$, with
\begin{equation}\label{eq5}
R=-2fhr^{2}h^{''}+fr^{2}h^{'}{}^{2}-rhh^{'}(rf^{'}+4f)+4h^{2}(-rf^{'}+k-f)=0,
\end{equation} 
where prime denotes derivative to $r$. Moreover, the $rr$ component of the field equations provides us with the second-order differential equations for $ f(r) $ and $h(r)$
\begin{align}\label{eq6}
0 &= E_{r r}= 6\alpha\left( -rh^{'} +2h\right)h^{2}f r^{2}f^{''}+3\alpha\Big(4kh^2-f r^{2}h^{\prime 2}-4fh^{2}-2rfh h^{\prime}\Big)h r f^{'}-9\alpha h^{3}{r}^{2}f^{'}{}^{2}\nonumber\\
&~~~~~~~~~+6\left(hr^{2}+r^{3}h^{\prime}+4\alpha k h\right)h^2f -3\alpha\left(3 h{r}^{2}h^{'}{}^{2}+8 h^{3}+ {r}^{3}h^{'}{}^{3}\right)f^{2}-6\,k r^{2}h^{3}.
\end{align}
Now, we are considering the near-horizon behaviors of the functions $h(r)$ and $f(r)$. By expanding $f(r)$ and $h(r)$ around the event horizon $({r_+})$, these read
\begin{align}\label{eq7}
h(r) &= h_{1}(r-r_{+})+h_{2}(r-r_{+})^{2}+h_{3}(r-r_{+})^{3}+..., \\
f(r)  &= f_{1}(r-r_{+})+f_{2}(r-r_{+})^{2}+f_{3}(r-r_{+})^{3}+...
\label{eq8}.
\end{align}
The coefficients $h_i$ and $f_i$ can be determined by inserting these expansions into \eqref{eq5} and $\eqref{eq6}$. Thus, we find
\begin{equation}\label{eq9}
h_{2}=\dfrac{h_{1}(k-2f_{1}r_{+})}{f_{1}r_{+}^{2}}+\dfrac{h_{1}(k- f_{1}r_{+})}{8\alpha f_{1}^{2}r_{+}},\;\;\;\;\;\;f_{2}=\dfrac{k-2f_{1}r_{+}}{r_{+}^{2}}-\dfrac{3(k-f_{1}r_{+})}{8\alpha f_{1}r_{+}},
\end{equation}
{where ${r_+}$, $h_1$ and $f_1$} are undetermined constants of integration. For a full expression of $h_3$ and $f_3$, see Appendix~\ref{sec:Appendix}.

We wish to obtain an approximate analytical solution  (for $k=1$) that is valid near the horizon and at large $r$.  To this end, we employ a continued fraction expansion by letting \cite{Rezzolla:2014mua}
\begin{equation}\label{eqq11}
h(r)=xA(x),\hspace{0.5cm}\dfrac{h(r)}{f(r)}=B^{2}(x),
\end{equation}
with
\begin{align}
A(x) &=1-\epsilon(1-x)+(a_{0}-\epsilon)(1-x)^{2}+\tilde{A}(x)(1-x)^{3},
\label{Ax}
\\
B(x) &=1+b_{0}(1-x)+\tilde{B}(x)(1-x)^{2},
\label{Bx}
\end{align} 
where
\begin{equation}
x = 1- \frac{r_+}{r}, \qquad 
\tilde{A}(x)=\dfrac{a_{1}}{1+\dfrac{a_{2}x}{1+\dfrac{a_{3}x}{1+\dfrac{a_{4}x}{1+...}}}},
\qquad 
\tilde{B}(x)=\dfrac{b_{1}}{1+\dfrac{b_{2}x}{1+\dfrac{b_{3}x}{1+\dfrac{b_{4}x}{1+...}}}}.
\label{cfrac}
\end{equation}
Note that, we truncate the continued fraction at order four. The coefficients $a_i$ and $b_i$ are given in terms of $({r_+}, h_1, f_1)$ by expanding \eqref{eqq11} near the horizon ($x\to 0$). We provide these expressions in the  Appendix~\ref{sec:Appendix}. 

Moreover, at large $r$, we assume
\begin{align}
h(r) = H_0 + \frac{H_1}{r} + \frac{H_2}{r^2} + ...,\;\;\;\;\;
f(r) = F_0 + \frac{F_1}{r} + \frac{F_2}{r^2} + ...,
\end{align}
where $H_i$ and $F_i$ are undetermined constants. By expanding \eqref{eqq11} near the asymptotic region ($ x\to 1 $), we obtain the following relations, 
% where we truncate the continued fraction at order $4$.
% By expanding (\ref{eq10}) near the horizon ($ x\to 0 $) and 
% the asymptotic  region ($ x\to 1 $)  we obtain  
\be
\epsilon=-\dfrac{H_{1}}{r_+}-1,  \qquad b_{0}=\dfrac{F_1-H_1}{2r_{+}},  \qquad a_{0}=\dfrac{H_{2}}{r_{+}^{2}}
\ee
{for the lowest order expansion coefficients. We find that $H_0=F_0=1$ and $H_2=0$. 
%\fixme{Are these from asymptotic flatness H=0 and F=0? If yes, does it mean we set $\Lambda=0$?} 
Moreover, we require $H_1=F_1=-2M$ (hence $b_0=0$) such that $h$ and $f$ only differ from $1/r^2$ term onward at the asymptotic region. A constant $M$ can be interpreted as a black hole's mass. 
% for the lowest order expansion coefficients, with the remaining
% $a_i$ and $b_i$ given in terms of $({r_+}, h_1, f_1)$; we provide these expressions in the  Appendix. 
In \cite{Kokkotas:2017zwt}, the coefficients of continued fraction expansion are obtained numerically and represented analytically and we have provided them in Appendix \ref{appE}. The major difference is that we obtain our coefficients analytically. 

For a static space-time, we have a timelike Killing vector 
$ \xi=\partial_{t} $ everywhere outside the horizon. Thus, we may define the temperature of a black hole as follows
\begin{align}\label{eq20}
T =&\dfrac{1}{4\pi}\left. \sqrt{\dfrac{f(r_{+})}{h(r_{+})}}h^{'}(r)\right\vert_{r_+}
= \dfrac{\sqrt{f_{1}h_{1}}}{4\pi}.
\end{align}
We compute the entropy by using the first two terms in continued fraction expansion as follows \cite{Wald1, Wald2, PhysRevD.92.124019}
\begin{align} 
S &=-2\pi\int_{Horizon}d^{2}x\sqrt{\eta}\dfrac{\delta L}{\delta R_{a b c d}}\epsilon_{a b}\epsilon_{c d},\nonumber \\
&=\dfrac{A}{4}\left[1+\dfrac{4\alpha (b_{1}^2+2 b_{1}+2\epsilon -a_{0}-a_{1})}{r_{+}^2(1+b_{1})^2}\right], \nonumber\\
&=\pi r_{+}^2\left[1+4\alpha\left( \dfrac{1}{r_{+}^{2}}-\dfrac{f^{'}({r_+})}{r_+}\right)\right], \nonumber \\
&=\pi r_{+}^2+4\pi \alpha \left(1 - f_{1}r_{+}\right),
\label{eq12}
\end{align}
here $\epsilon_{ab}$ is the binormal tensor to the black hole horizon, normalized as $\epsilon_{ab}\epsilon^{ab}=-2$ and $\eta$ is determinant of the induced metric {on} the horizon.
We consider the thermodynamics of these black hole solutions, whose basic equations are 
the first law and Smarr formula 
\begin{equation}\label{eqfirstlaw}
dM=TdS,\;\;\;\;\;\;\;\; M=2TS.
\end{equation}
It should also be noted that the effect of higher curvature gravity in the first law of thermodynamics and the Smarr formula is not taken into account. Because in \cite{Hajian:2023bhq}, it has been shown that the coupling constant of a theory is a thermodynamic quantity if the {on-shell} Lagrangian of the higher curvature gravity satisfies the conditions (11) of \cite{Hajian:2023bhq}, i.e. $L_{QG}=dA$, here $A$ is a gauge field. For quadratic gravity, the mentioned condition is not satisfied.
%\fixme{Please add heat capacity and free energy here}
In the following subsection{s}, we will consider the thermodynamic stability of the novel solutions. For global stability, we allow a system in equilibrium with a thermodynamic reservoir to exchange energy with the reservoir. The preferred phase of the system is the one that minimizes the free energy $(F)$. To investigate the global stability, we use the following expression for the free energy
\begin{equation}
F=M-TS.
\end{equation}
On the other hand, local stability is concerned with how the system responds to small changes in thermodynamic parameters. To study the thermodynamic stability of the black holes concerning small variations of the thermodynamic coordinates, one can investigate the behavior of the heat capacity. The heat capacity is given by 

\begin{equation}
C=\dfrac{\partial M}{\partial T}.
\end{equation}
The positivity of the heat capacity ensures local stability. {In the next {sub}sections \ref{linearcase}, we review analytic solutions in the case of $f_{1}=h_{1}$ according to \cite{Sajadi:2020axg} and compare the results with the numerical solutions reported in \cite{Lu:2015cqa,Bonanno:2019rsq,Kokkotas:2017zwt}.}

\subsection{{The case $h_{1}=f_{1}$}}\label{linearcase}

Here, we consider the case of $h_{1}=f_{1}$. The computations regarding this case with details provided in \cite{Sajadi:2020axg}. At first, we obtain a relation between physical parameters of theory as follows:
\begin{equation}\label{c1pplot}
c_{1}=%-\dfrac{12\alpha r_{+}^4+r_{+}^6+64\alpha^3 +48\alpha^2 r_{+}^2}{432\alpha^2 r_{+}^2}=
-\dfrac{0.009259259256(p^2+2)^3}{p^2},
\end{equation}  
here $p=r_{+}/\sqrt{2\alpha}$ is a dimensionless parameter. The above condition has been obtained from the intersection between the non-Schwarzschild-like and Schwarzschild-like branches. In figure \ref{c1rplotta1}, using the above equation the behavior of constant of integration $c_{1}$ and $f_{1}$ have been shown. As can be seen, by increasing $c_{1}$ in the left panel of figure \ref{c1rplotta1}, the non-Schwarzschild and Schwarzschild branches get closer to each other ({red solid lines}). At $p=p_{s}=1$ and $c_1=c_{1s}=-0.25$, the non-Schwarzschild and Schwarzschild black holes coalesce (black dot lines). 
Then, by increasing $c_{1}$, the non-Schwarzschild and Schwarzschild branches are separated again ({blue solid line}). Here, the point is that below $c_{1}<c_{1s}$, the small Schwarzschild, non-Schwarzschild black hole and large Schwarzschild, non-Schwarzschild black hole exist separately. 

For $c_{1}>c_{1s}$,  the small Schwarzschild and large non-Schwarzschild black holes {coexist}, likewise the large Schwarzschild and small non-Schwarzschild black holes {coexist}. In contrast to the numerical work \cite{Lu:2015cqa, Bonanno:2019rsq, Kokkotas:2017zwt}, only for $c_1=c_{1s}$ and $p=p_{s}$ both the Schwarzschild and non-Schwarzschild black holes are the exact solutions of the Einstein-Weyl gravity. 

Notice that, numerically for all $p$ the Schwarzschild metric is the solution of the Einstein-Weyl equations, but at some minimal nonzero $p_{min}$ in addition to the Schwarzschild solution, there appears the non-Schwarzschild branch \cite{Lu:2015cqa, Kokkotas:2017zwt}. 

According to our results obtained for $f_{1}$ and $M$, we have shown the behavior of the thermodynamic quantities i.e., mass and entropy in figure \ref{mrplotta1}. We compare our analytic results with those obtained in \cite{Sajadi:2020axg} (solid red lines) and numerical results obtained in \cite{Lu:2015cqa,Bonanno:2019rsq,Kokkotas:2017zwt} (dot green lines). {We remark that the Schwarzschild-like branch is represented by the solid line in which the black hole's mass increases with ${r_+}$. On the other hand, the non-Schwarzschild-like solution is the solid curve in the black hole's mass decreases with ${r_+}$. }

As can be seen, in each figure the thermodynamic quantities are the same as the numerical ones for the Schwarzschild-like branches. However, for the non-Schwarzschild-like branches, the analytical and numerical quantities are significantly different. This agrees with the results of the paper \cite{Bonanno:2019rsq} in which $f_{1}=h_{1}$ gives the results of Einstein's gravity. However, in this case, the important point is that there are extra branches in the analytical thermodynamics quantities that have different behavior from the Schwarzschild branches. These branches have behavior similar to the non-Schwarzschild-like branches in numerical results \cite{Lu:2015cqa} i.e., the mass of the black hole decreases with $r_{+}$ and becomes zero at a certain value of $r_{+}$. These {new} branches are obtained for every value of the coupling constant of the theory ($\alpha$). Note that, in figure \ref{mrplotta1}, we fix $\alpha=0.5$ to compare with the numerical solutions obtained in \cite{Lu:2015cqa,Bonanno:2019rsq,Kokkotas:2017zwt}.

\begin{figure}[H]
\centering
\includegraphics[width=0.45\columnwidth]{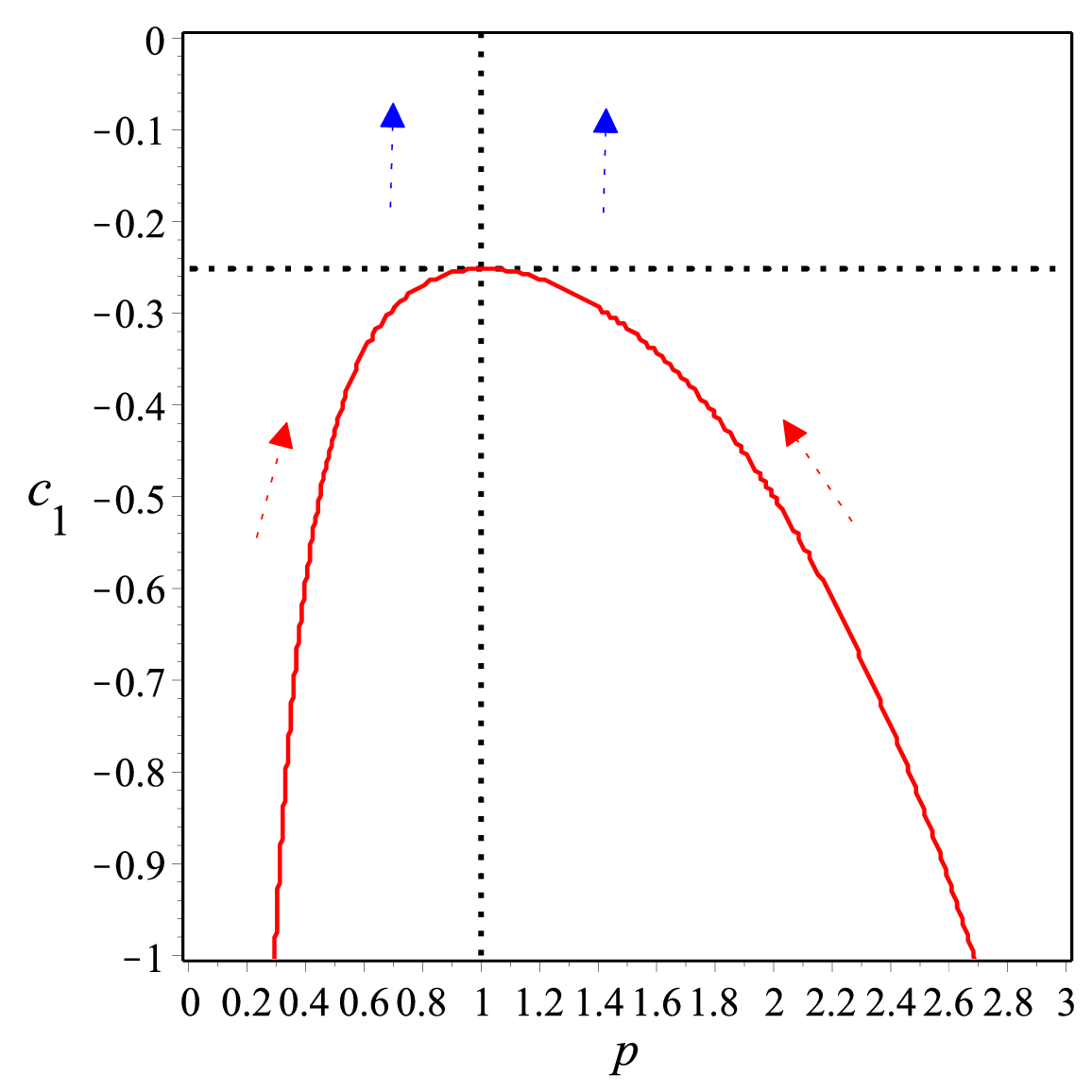}
\includegraphics[width=0.45\columnwidth]{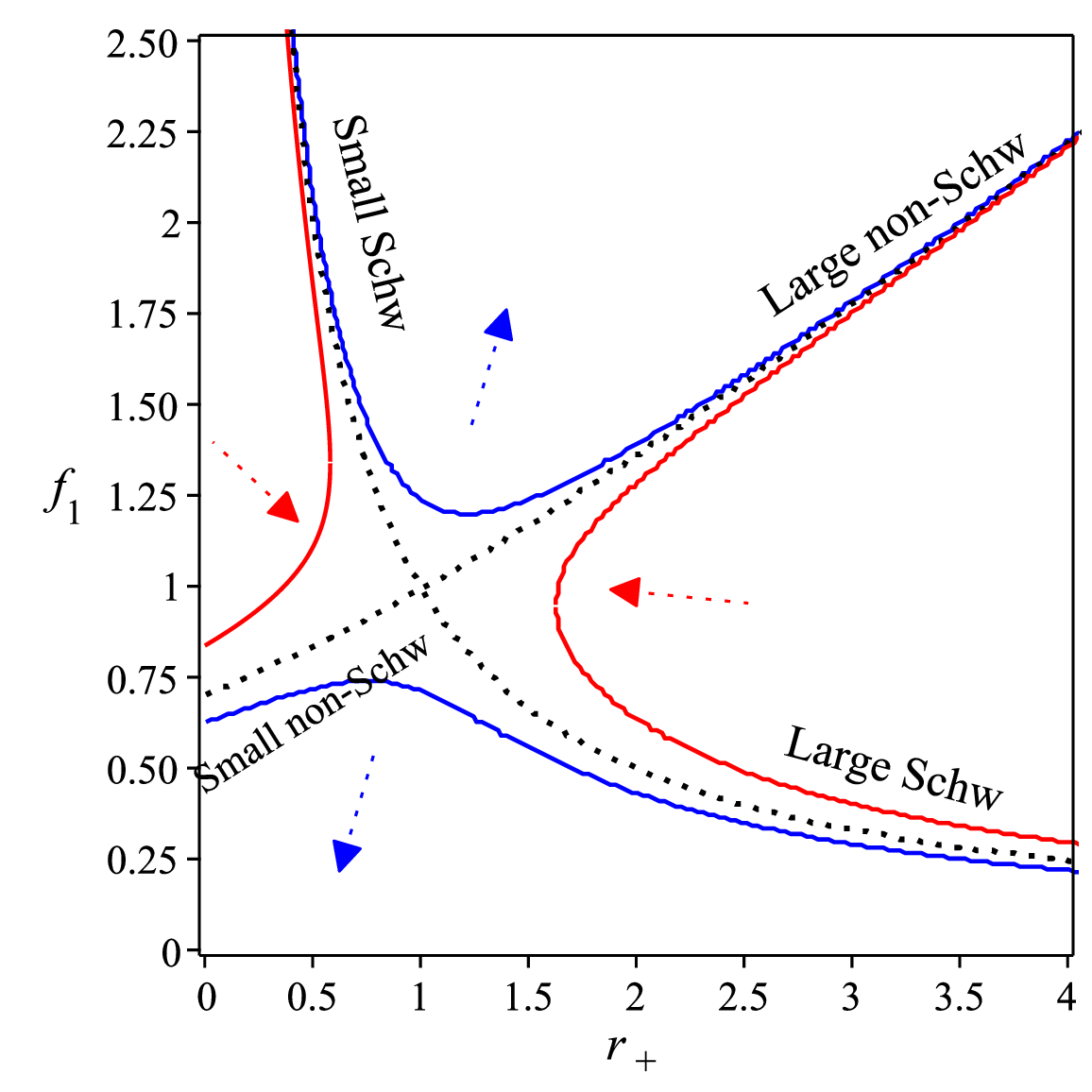}
\caption{\small
The behavior of $c_{1}$ in terms of $p$ (left). The behavior of $f_{1}$ in terms of $r_{+}$ for $\alpha=0.5$ and $c_{1}=\textcolor{red}{-0.35},-0.25,\textcolor{blue}{-0.2}$ (right).}
\label{c1rplotta1}
\end{figure}

In figure \ref{fhhfplotta1}, based on the results obtained for $f_{1}$ and $M$, we have shown the behavior of metric functions for the non-Schwarzschild-like branch (blue curves) comparing with the numerical one (red curves) in \cite{Kokkotas:2017zwt}. In the right panel of figure \ref{fhhfplotta1}, we can see that there is a good agreement between our analytical results and the numerical one, especially at small and large $r$. For the largest difference occurs around $r\approx 3.5$ where the maximum difference is less than 10$\%$. We remark that here our obtained solution is constructed up to $a_4$ in the continued fraction expansion \eqref{cfrac}. By including higher coefficients, it is possible to get more accurate metric functions.

\begin{figure}[H]
\centering
\includegraphics[width=0.45\columnwidth]{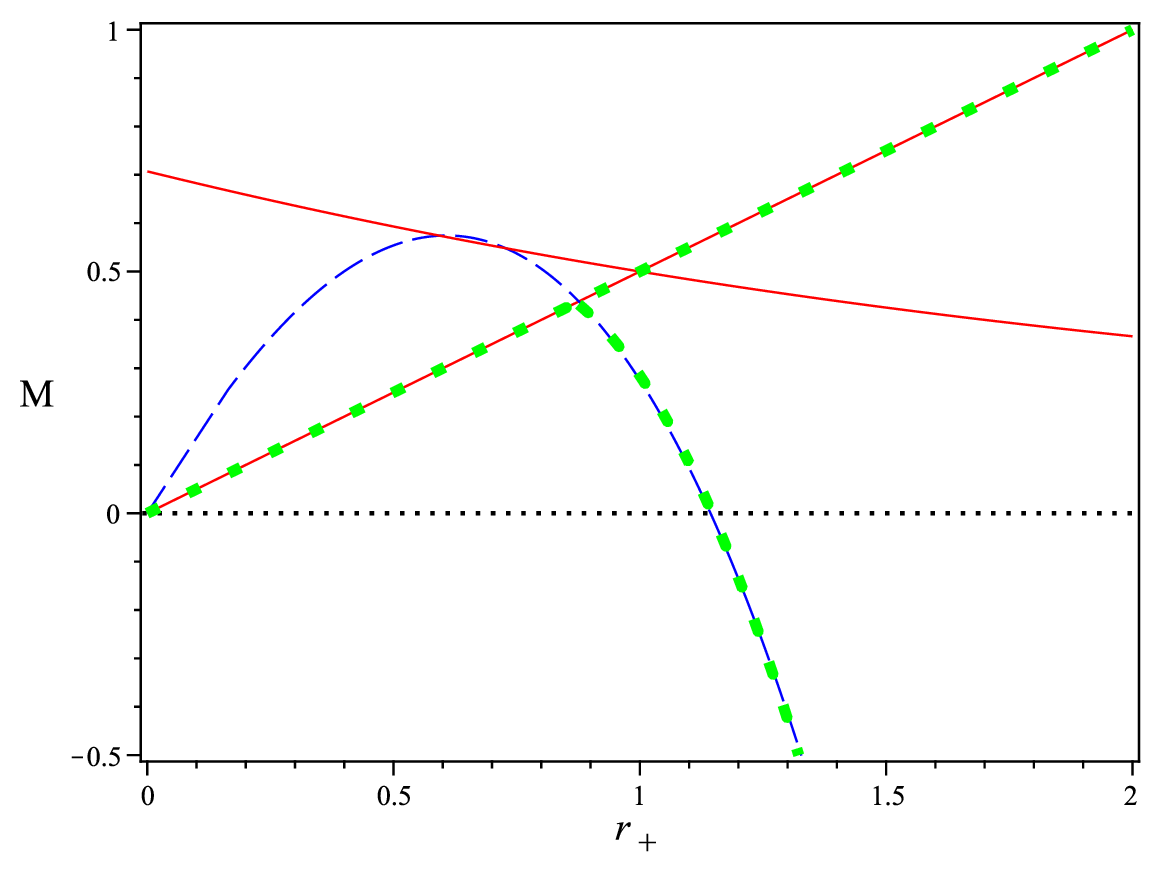}
\includegraphics[width=0.49\columnwidth]{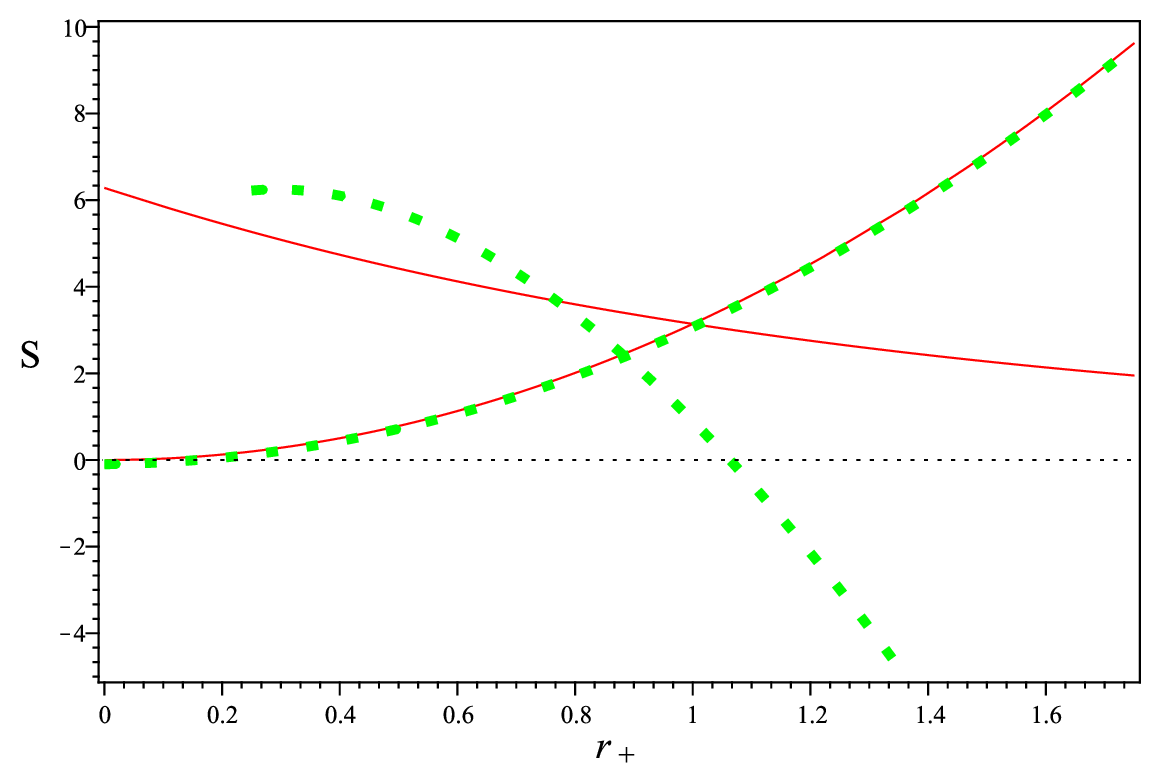}
\caption{\small
The behavior of mass (left) and entropy (right) in terms of $r_{+}$ for $\alpha=0.5, c_{1}=-0.25$. The dot green lines are the numerical results \cite{Lu:2015cqa}; the blue dashed line is the numerical result according to \cite{Kokkotas:2017zwt} and the solid red lines are our results.}
\label{mrplotta1}
\end{figure}

\begin{figure} 
\centering
\includegraphics[width=0.45\columnwidth]{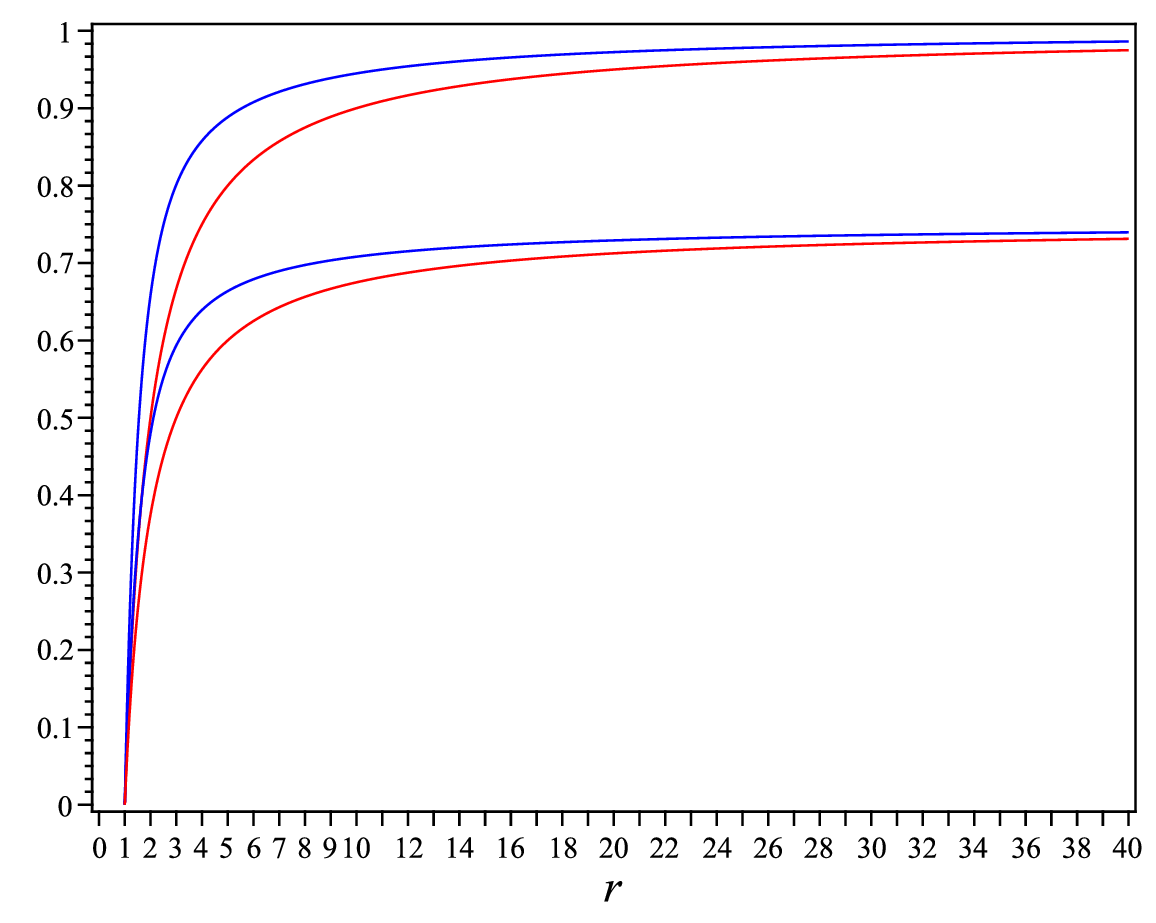}
\includegraphics[width=0.45\columnwidth]{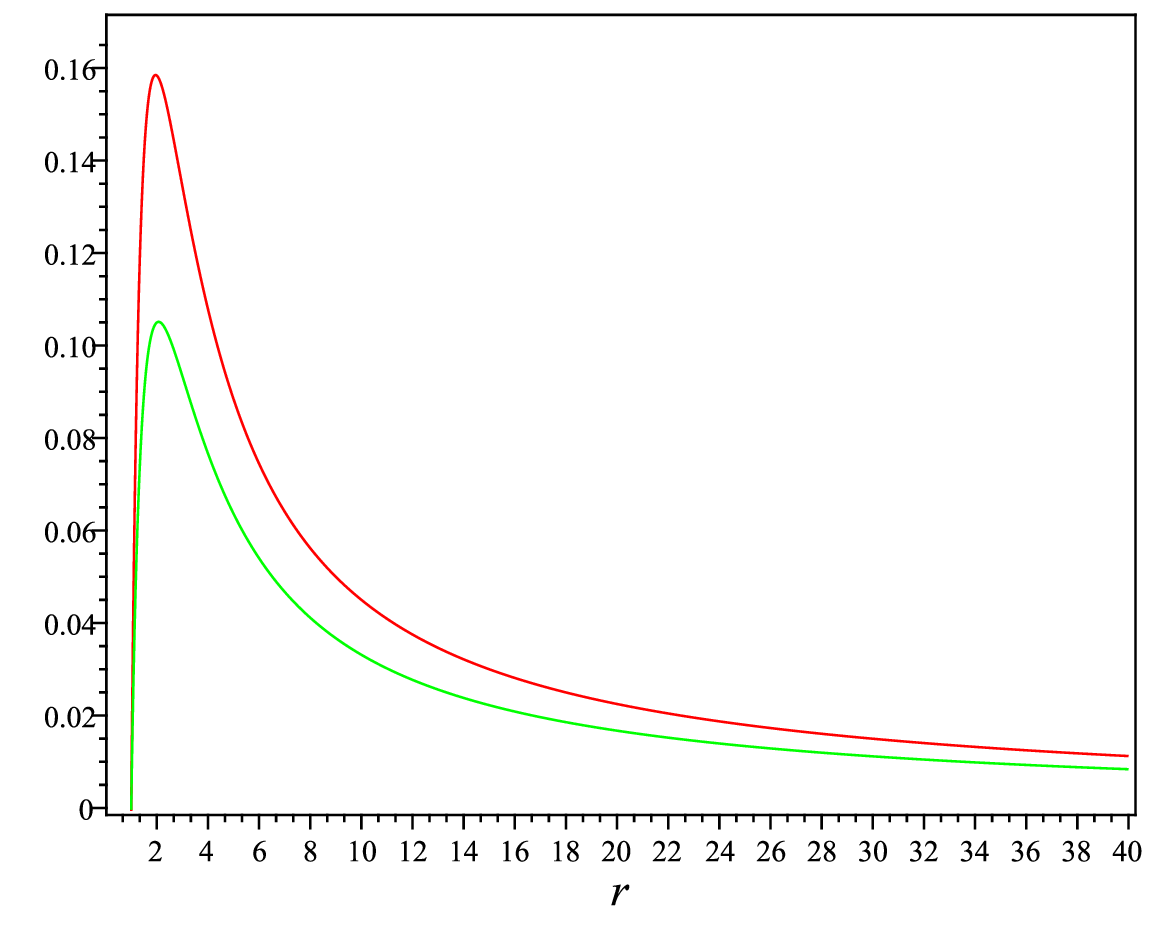}
\caption{\small Left: The behavior of $h(r)$ and $0.75f(r)$ (\textcolor{blue}{blue lines}, upper) according to our analytic results in non-Schwarzschild-like branch; The numerical results for $h(r)$ and $0.75f(r)$ (\textcolor{red}{red lines}) according to \cite{Kokkotas:2017zwt} (lower). Right: The difference between analytical and numerical approximations for $h(r)$ (\textcolor{red}{red line}, upper) and $f(r)$ (\textcolor{green}{green line}, lower).}
\label{fhhfplotta1}
\end{figure}

In this subsection, we find that for the case of a linear relation between near-horizon coefficients i.e., $f_1=h_1$, the analytical and numerical results of the Schwarzschild branch are the same. In the following subsection, we shall consider the case $f_1 \neq h_1$ where it is suggested in \cite{Bonanno:2019rsq} that this case corresponds to the non-Schwarzschild BHs of EQG.

\subsection{The case $h_{1}=f_{1}^2$}\label{secc3}
We now consider the following relation between the near horizon parameters $f_1$ and $h_1$:
\begin{equation}
h_1({r_+})=f_1^2(r_{+}).
\end{equation}
The above relation is approximately similar to the result in \cite{Bonanno:2019rsq} (figure 2). With the above relation, one finds from equations (\ref{eqfirstlaw})
\begin{align}
&\;\;\;\;\;\;\;\;\;\;\;\;\;\;\;\;\;\;\;\;\;\;\;\;\;M=-\dfrac{1}{2}f^{\frac{3}{2}}_{1}(r_{+})\left(4\alpha r_{+}f_{1}(r_{+})-r_{+}^{2}-4\alpha \right),\label{eqqmass19}\\
&\dfrac{3}{4}\sqrt{f_{1}(r_{+})}\left[f^{\prime}_{1}(r_{+})\left(-\dfrac{16}{3}\alpha r_{+}f_{1}(r_{+})+r_{+}^{2}+4\alpha\right)+\dfrac{2}{3}f_{1}(r_{+})\left(r_{+}-2\alpha f_{1}(r_{+})\right)\right]=0.\label{eqq20}
\end{align}
Solving \eqref{eqq20}, one obtains a quartic equation as follows
\begin{equation}
4f^{4}_{1}(r_{+})r_{+}\alpha -(r_{+}^{2}+4\alpha)f^{3}_{1}(r_{+})+c_{1}=0,
\end{equation}
where $c_1$ is the integration constant. This leads to four solutions for $f_{1}(r_{+})$ as follows
\begin{align}\label{eqq2324}
f_{1}^{1,2}(r_{+})=&\dfrac{\pm\sqrt{6\mathcal{C}\sqrt{\mathcal{B}}+18\sqrt{3\mathcal{A}}(r_{+}^{2}+4\alpha)^{3}}+\sqrt{3}\mathcal{B}^{\frac{3}{4}}+3(r_{+}^2+4\alpha )\mathcal{A}^{\frac{1}{6}}\mathcal{B}^{\frac{1}{4}}}{48\alpha r_{+}\mathcal{A}^{\frac{1}{6}}\mathcal{B}^{\frac{1}{4}}},\\
f_{1}^{3,4}(r_{+})=&\dfrac{\pm\sqrt{\mathcal{D}\sqrt{\mathcal{E}}-18\sqrt{3\mathcal{A}}(r_{+}^{2}+4\alpha)^{3}}-\sqrt{3}\mathcal{E}^{\frac{3}{4}}+3(r_{+}^2+4\alpha )\mathcal{A}^{\frac{1}{6}}\mathcal{E}^{\frac{1}{4}}}{48\alpha r_{+}\mathcal{A}^{\frac{1}{6}}\mathcal{E}^{\frac{1}{4}}},
\end{align}
with
\begin{align}\label{eqq24}
\mathcal{A} &= 9r_{+}^{4}+72\alpha r_{+}^{2}+144\alpha^2 +\left(-49152 c_{1}r_{+}^{3}\alpha^{3}+81r_{+}^{8} +1296\alpha r_{+}^{6}+7776\alpha^{2}r_{+}^{4} \right. \nonumber \\
&~~~\left. +20736\alpha^3 r_{+}^{2}+20736 \alpha^4\right)^{\frac{1}{2}},\nonumber \\
\mathcal{B} &= 128\alpha^{2}r_{+}^{2}(12c_{1})^{\frac{2}{3}}+8\alpha r_{+}(12 c_{1})^{\frac{1}{3}}\mathcal{A}^{\frac{2}{3}}+3(r_{+}^2+4\alpha)^{2}\mathcal{A}^{\frac{1}{3}},\nonumber\\
\mathcal{C} &= -64\alpha^2 r_{+}^{2}(12c_{1})^{\frac{2}{3}}-4\alpha r_{+}(12c_{1})^{\frac{1}{3}}\mathcal{A}^{\frac{2}{3}}+3(r_{+}^2+4\alpha)^{2}\mathcal{A}^{\frac{1}{3}},\nonumber\\
\mathcal{D} &= -768\alpha^2 r_{+}^2 (12c_{1})^{\frac{2}{3}}-24\alpha r_{+}(3c_{1})^{\frac{1}{3}}(2\mathcal{A})^{\frac{2}{3}}+18(r_{+}^{2}+4\alpha)^{2}\mathcal{A}^{\frac{1}{3}},\nonumber\\
\mathcal{E} &= 256\alpha^{2}r_{+}^{2}(12c_{1})^{\frac{2}{3}}+8\alpha r_{+}(3c_{1})^{\frac{1}{3}}(2\mathcal{A})^{\frac{2}{3}}+3(r_{+}^{2}+4\alpha)^{2}\mathcal{A}^{\frac{1}{3}}.
\end{align}
There are multiple branches of analytic solution regarding, $f_1^{1,2,3,4}$ given that they are real-valued constants. 
By inserting them into \eqref{eq20}, \eqref{eq12}, and \eqref{eqqmass19} one can determine the thermodynamics quantities as a function of ${r_+}$. Before that, similar to the previous subsection,
we can obtain a relation between the parameters of the theory as follows:
\begin{equation}
c_{1}=%\dfrac{27(r_{+}^{8}+16\alpha r_{+}^{6}+96\alpha^2 r_{+}^{4}+256\alpha^3 r_{+}^2+256\alpha^4)}{16384\alpha^3 r_{+}^{3}}=
\dfrac{0.009322208539(p^2+2)^4}{\sqrt{\alpha}p^3},
\end{equation}
here $p=r_{+}/\sqrt{2\alpha}$. From figure \ref{c1f1plottasquar1}, the two branches of solutions get closer to each other as $c_1$ decreases. At $c_{1}=c_{1s}=1.05166$ and $p=p_{s}=1.1$, the two solutions intersect each other. For $c_{1}<c_{1s}$, the two solutions are separated. The same trend is observed for the behavior of thermodynamics quantities in the figures \ref{MSTrpplottasquar1}.

\begin{figure} 
\centering
\includegraphics[width=0.4\columnwidth]{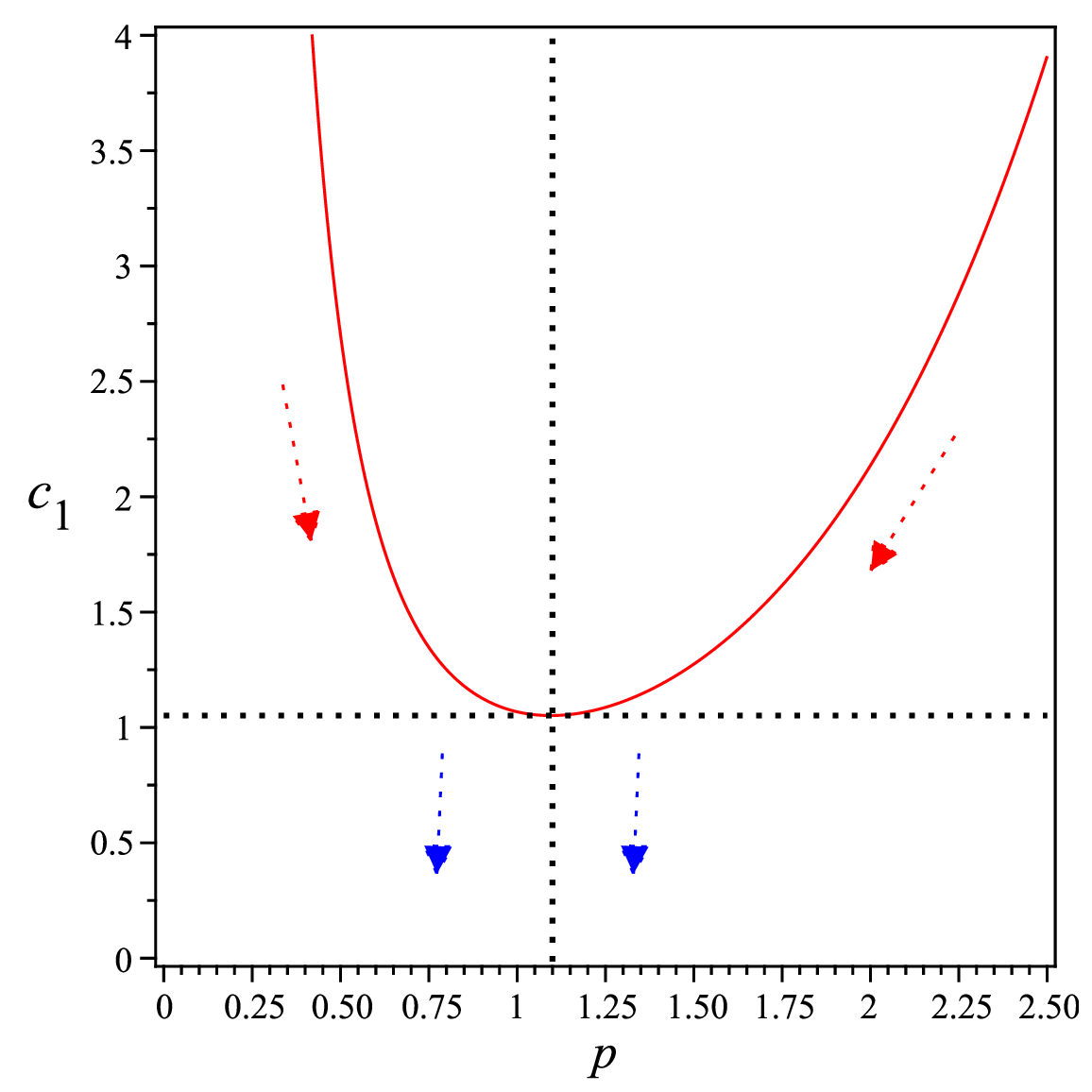}
\includegraphics[width=0.4\columnwidth]{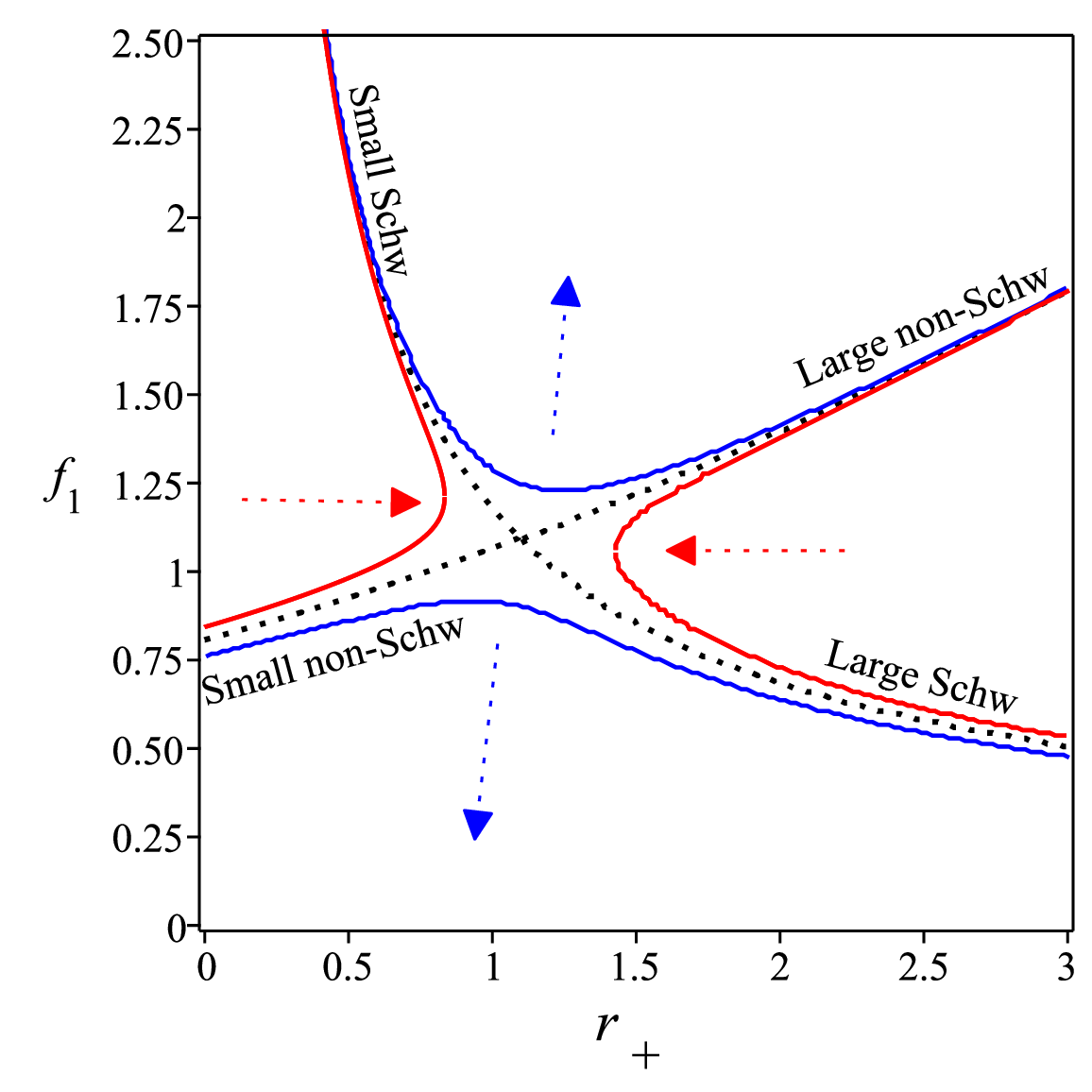}
\caption{\small The behavior of $c_{1}$ in terms of $p$ (left). The behavior of $f_{1}$ in terms of $r_{+}$ for $\alpha=0.5$ and $c_{1}=\textcolor{red}{1.2},1.05166,\textcolor{blue}{0.9}$ (right).}
\label{c1f1plottasquar1}
\end{figure}
We look at the critical case ($c_{1s}=1.05166,p_{s}=1.1$) with more details in figure \ref{MSTrpplottasquarcrit}.
As can be seen, there are two kinds of solutions with four branches {of $f_1({r_+})$} in which the black dashed line is the Schwarzschild-like solution and the solid line is a non-Schwarzschild-like solution. The solid colored line consists of two branches of solution i.e., $f_1=f_1^1$ (blue) and $f_1=f_1^2$ (red). For the non-Schwarzschild solution, $M$ and $S$ decrease with the event horizon while the temperature increases with ${r_+}$. The Schwarzschild-like solutions have the opposite trends. Another observation from this figure is that the solutions exist for arbitrary values of the $\alpha$  and a particular value of the constant of the integration $c_1$. 

In figure \ref{figmst1}, we depict the behavior of $S$ as a function of $M$ and $M$ as a function of $T$. We observe a small Schwarzschild-like behavior that starts with the red solid curve and then smoothly connects to a large Schwarzschild-like black hole which is the black solid line. While small non-Schwarzschild-like behavior first follows the blue solid line and then continues with the orange dashed line to a large non-Schwarzschild-like black hole.

In figure \ref{figmstcomb1}, we have compared the analytical thermodynamics quantities with the numerical ones. The analytical thermodynamics quantities of this case are different from the numerical ones for both branches ({Schwarzschild-like and non-Schwarzschild-like)}. We expect the analytical Schwarzschild branch to be different from its numerical one for $h_{1}\neq f_{1}$ \cite{Bonanno:2019rsq}. {The non-Schwarzschild branch although it is very different from the numerical one has become closer to it concerning linear relation $f_{1}=h_{1}$.}

In the last and perhaps most important step, by equating the coefficients of continued fraction expansion of Appendix \ref{sec:Appendix} with \ref{appE}, we obtain $f_{1}$ and $h_{1}$ in terms of $r_{+}$ as follows:
\begin{align}
h_{1}=&\dfrac{3704000354214r_{+}^{2}-2463952425415\sqrt{2\alpha}r_{+}+456461849800\alpha}{1825490389932\alpha r_{+}},\label{eqh131}\\
f_{1}=&\dfrac{215672004025\alpha(246395242515\sqrt{2\alpha}r_{+}-3704000354214r_{+}^2-456461849800\alpha)}{456372597483r_{+}A}\label{eqf132},
\end{align}
here
\begin{align}
A=&63690786865\sqrt{2}r_{+}\alpha^{\frac{3}{2}}+792505343352\sqrt{2\alpha}r_{+}^3-145428585201r_{+}^4-\nonumber\\
&2393109251108\alpha r_{+}^2-93928764484\alpha^2.
\end{align}
It is easy to obtain $r_{+}$ in terms of $h_{1}$ from equation \eqref{eqh131}, and by inserting it into the equation \eqref{eqf132}, we can obtain a relation between $h_{1}$ and $f_{1}$. It should be noted these relations only work for the large non-Schwarzschild-like branch (Because the coefficients of continued fraction expansion of Appendix \ref{appE} only work for a non-Schwarzschild branch.).
This is one of the new results of this paper. We find an analytical relationship for $f_{1}$ and $h_1$ using thermodynamics quantities. Our results also agree with the numerical one in the large non-Schwarzschild black hole.

\begin{figure}[h]\hspace{0.4cm}
\centering
%\subfigure{\includegraphics[width=0.45\columnwidth]{h1f1rpexactplot}}
\subfigure{\includegraphics[width=0.45\columnwidth]{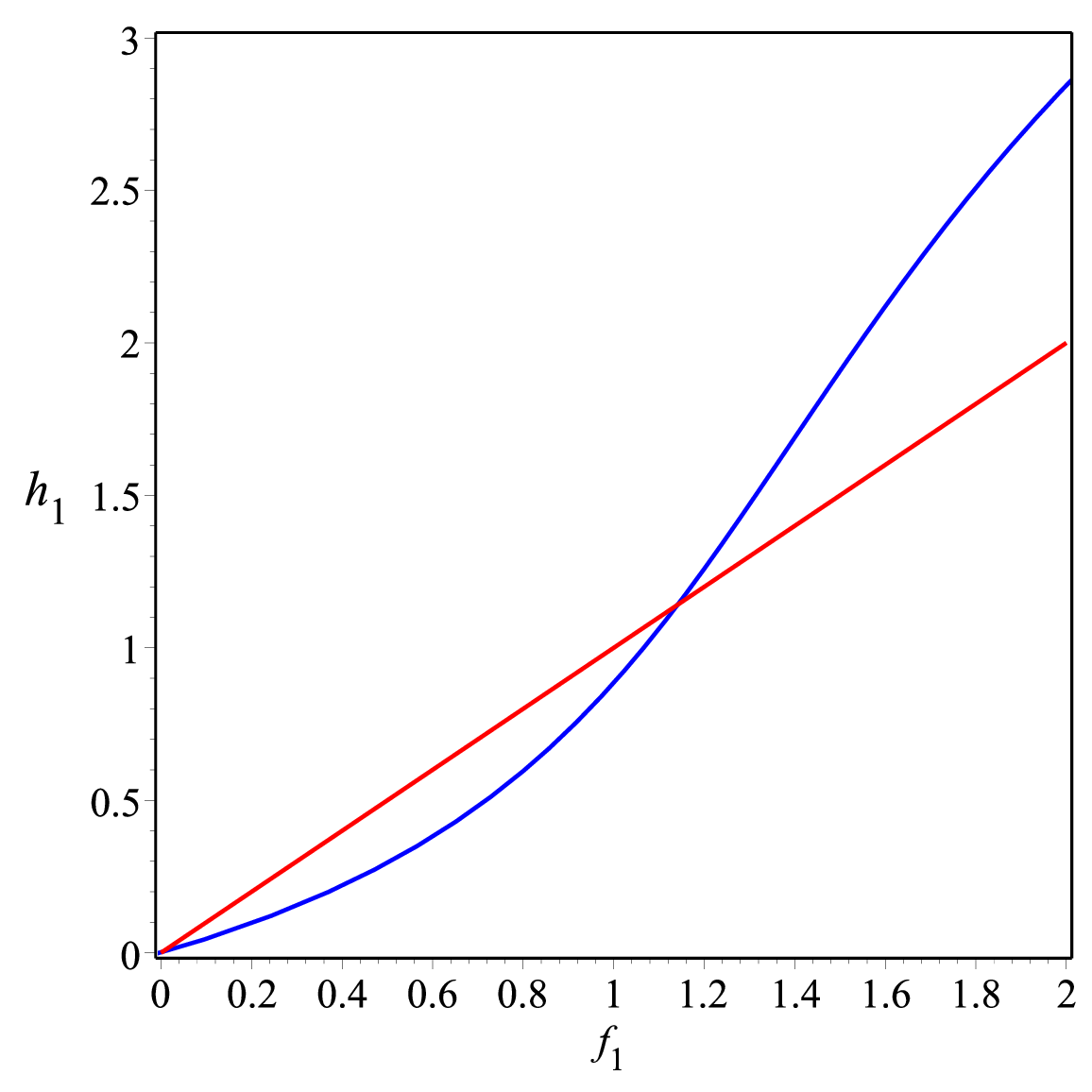}}
%\subfigure{\includegraphics[width=0.3\columnwidth]{FTTplot1}}
\caption{The behavior of $h_{1}$ in terms of $f_{1}$ for $ \alpha=0.5$.} 
\label{figf1h1rp}
\end{figure}

In figure \ref{figf1h1rp}, we have shown the behavior of $h_{1}$ vs $f_{1}$. The blue curve shows the behavior of non-Schwarzschild black holes. In figure \ref{figcoef}, the behavior of the coefficients of continued fraction expansion of Appendix \ref{sec:Appendix} and \ref{appE} from the inserting of \eqref{eqh131}, \eqref{eqf132} have been shown. As can be seen, there is a good agreement between the results.
\begin{figure}[h]\hspace{0.4cm}
\centering
\subfigure{\includegraphics[width=0.3\columnwidth]{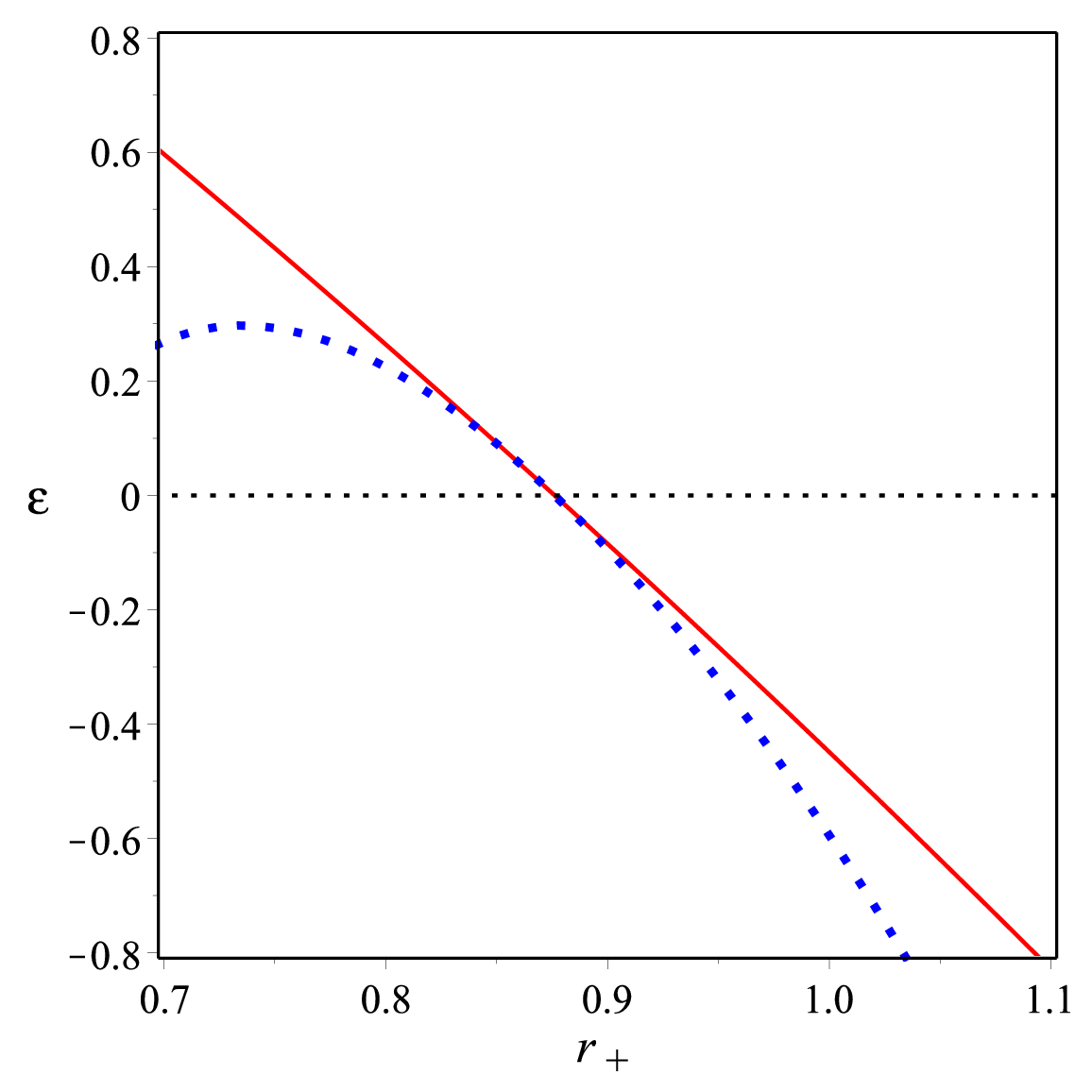}}
\subfigure{\includegraphics[width=0.3\columnwidth]{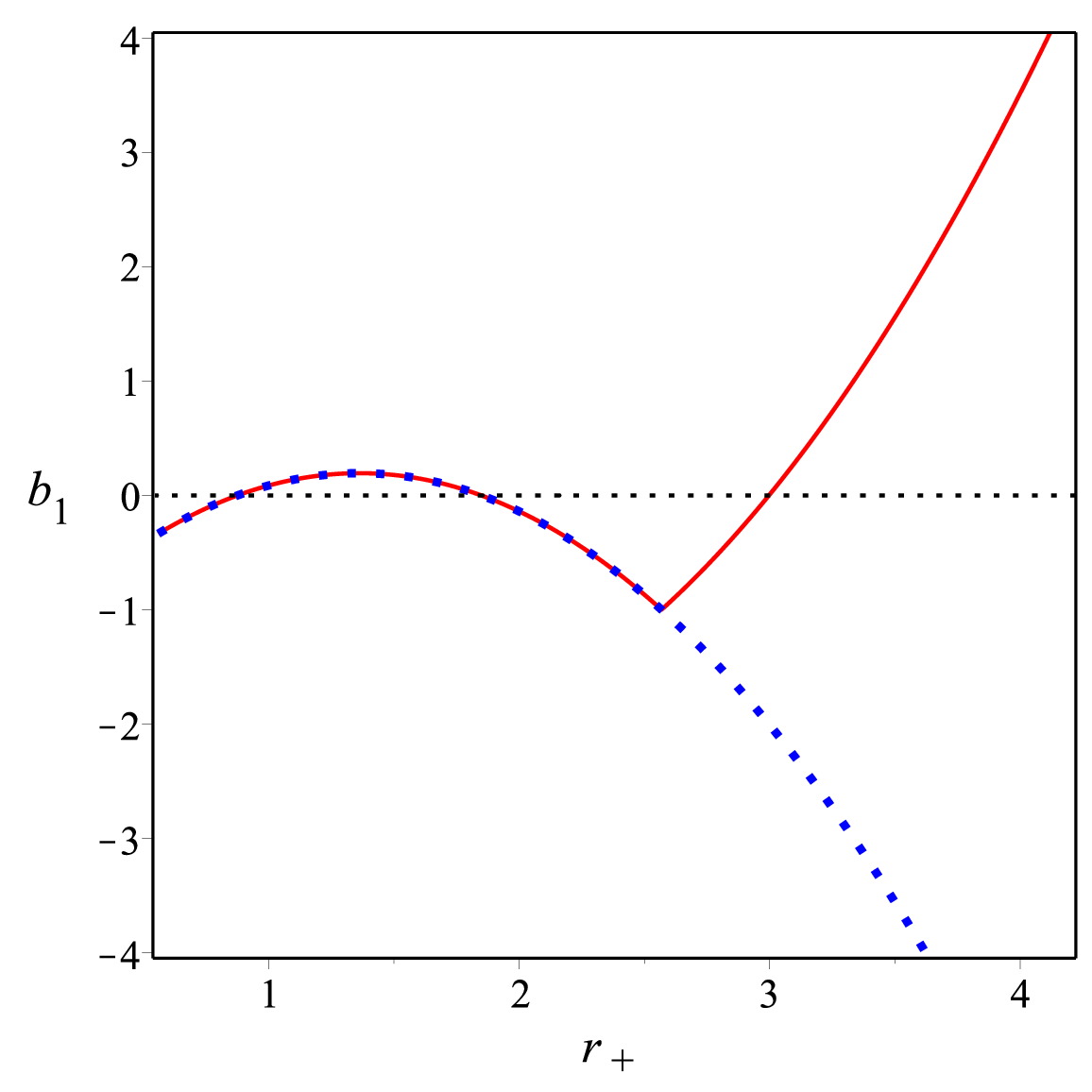}}
\subfigure{\includegraphics[width=0.3\columnwidth]{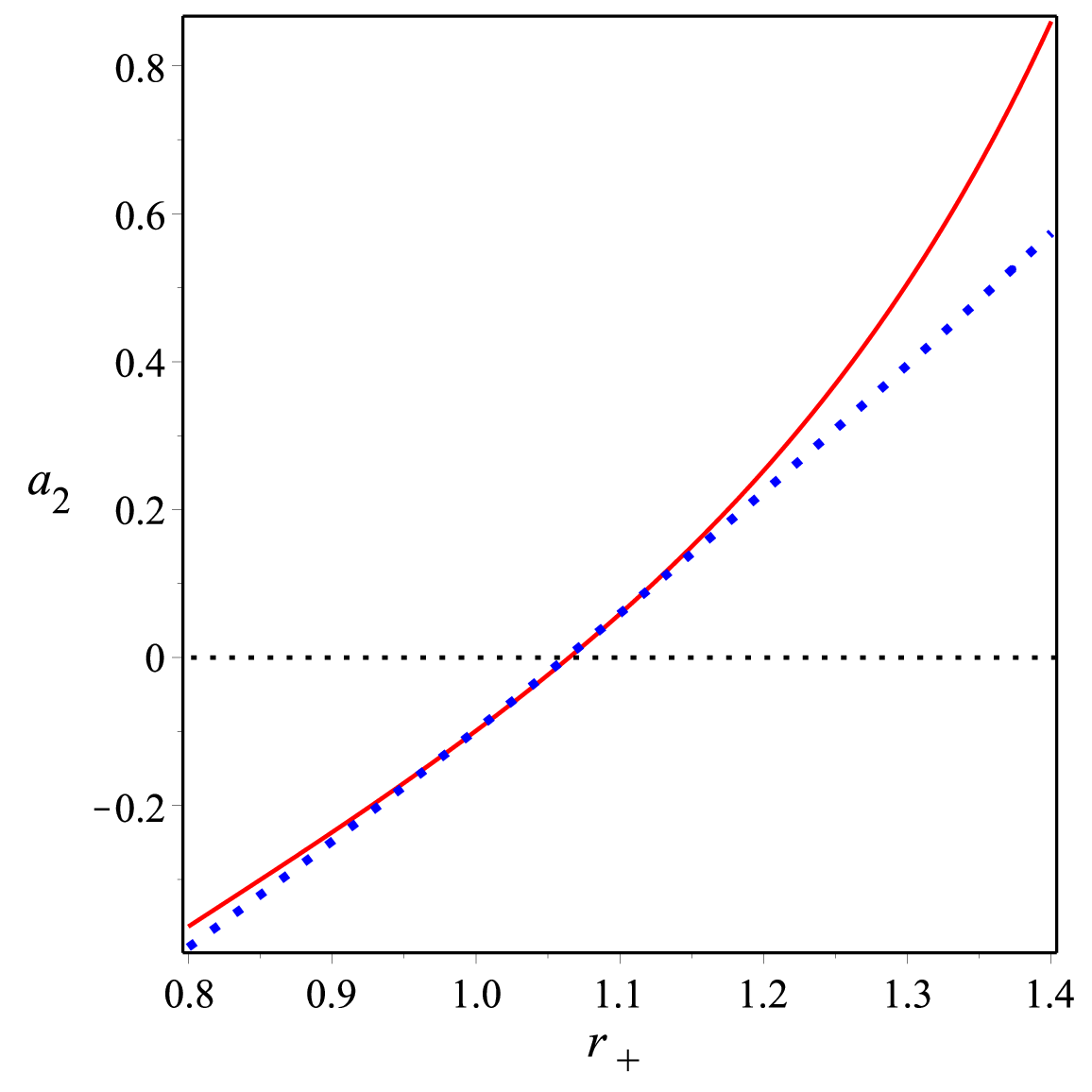}}
\caption{The behavior of coefficients of continued fraction expansion in terms of $r_{+}$ for $\alpha=0.5$. Dot lines result from \cite{Kokkotas:2017zwt} and solid lines are the results of our method.
} 
\label{figcoef}
\end{figure}

 In figure \ref{figmstcomb1} for the $f_{1}$ and $h_{1}$ in equations \eqref{eqh131}, \eqref{eqf132} the behavior of thermodynamics quantities have been shown in solid gold colors. As can be seen, they agree with the numerical thermodynamics quantities of non-Schwarzschild black holes within a good approximation. 

\begin{figure}[H]
\centering
\subfigure%[$ c_{1}=\textcolor{blue}{0.9},1.052,\textcolor{red}{1.2}$]
{
 \includegraphics[width=0.3\columnwidth]{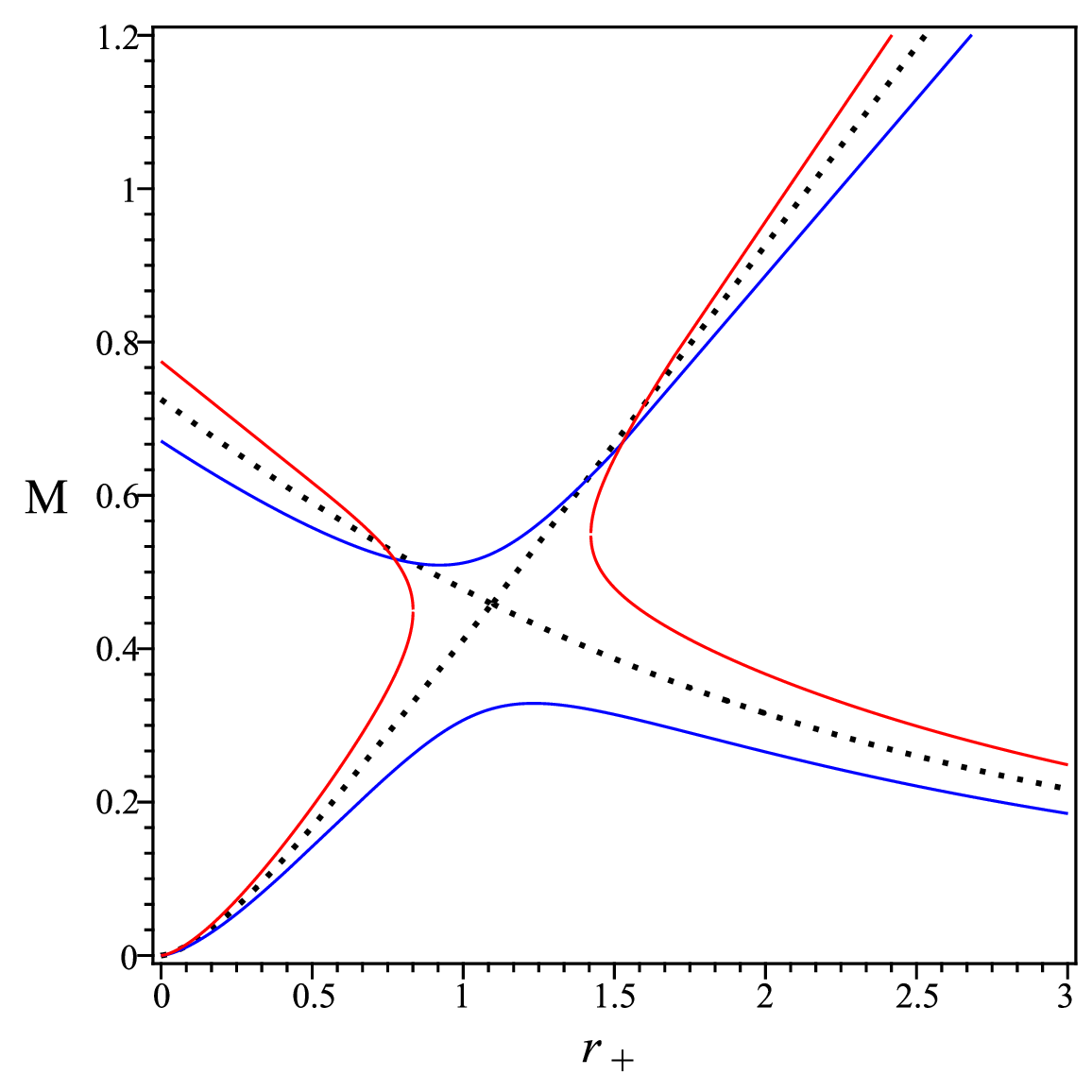}
% \label{fignT0}
 }
 \subfigure%[$ c_{1}=\textcolor{blue}{0.9},1.052,\textcolor{red}{1.2}$]
 {
 \includegraphics[width=0.3\columnwidth]{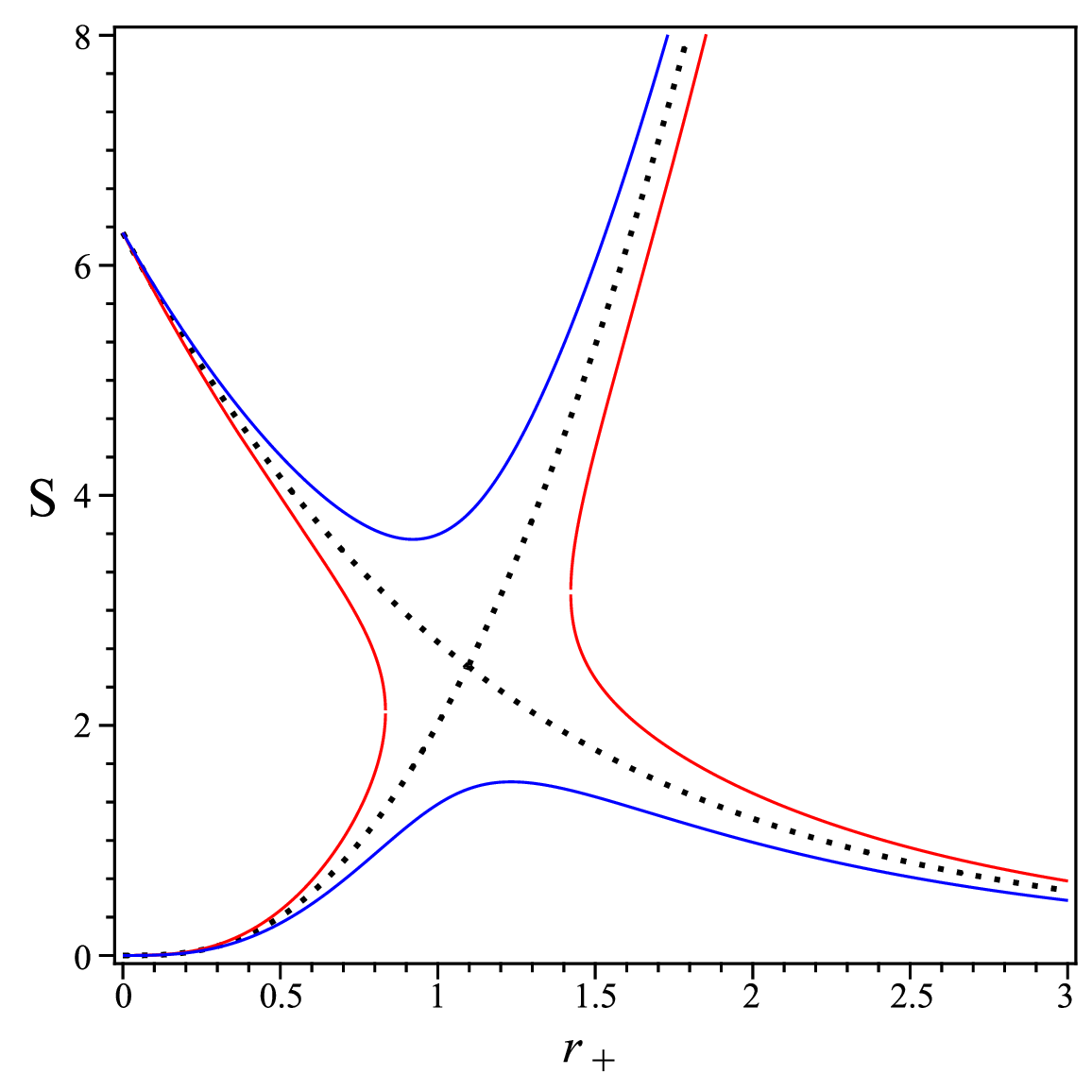}
 %\label{fignM0}
 }
 \subfigure
 {
 \includegraphics[width=0.3\columnwidth]{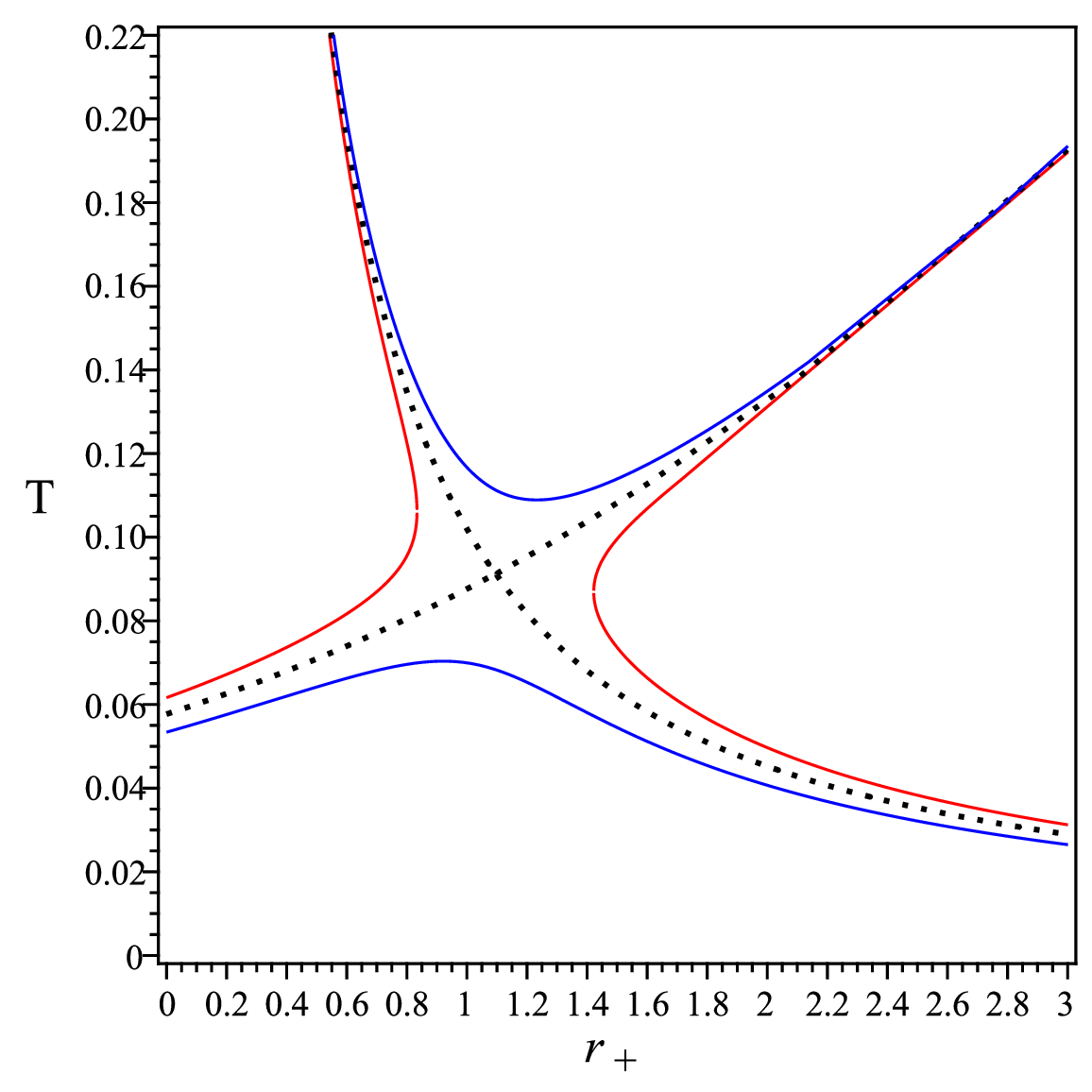}
 %\label{fignS0}
 }
 \caption{Plots of $ M $, $ S $ and $T$ in terms of $r_{+}$ for $\alpha=0.5, c_{1}=\textcolor{blue}{0.9},1.052,\textcolor{red}{1.2}$.}
 \label{MSTrpplottasquar1}
\end{figure}

\begin{figure}[H]
\centering
\subfigure%[$ c_{1}=1.052$]
{
 \includegraphics[width=0.3\columnwidth]{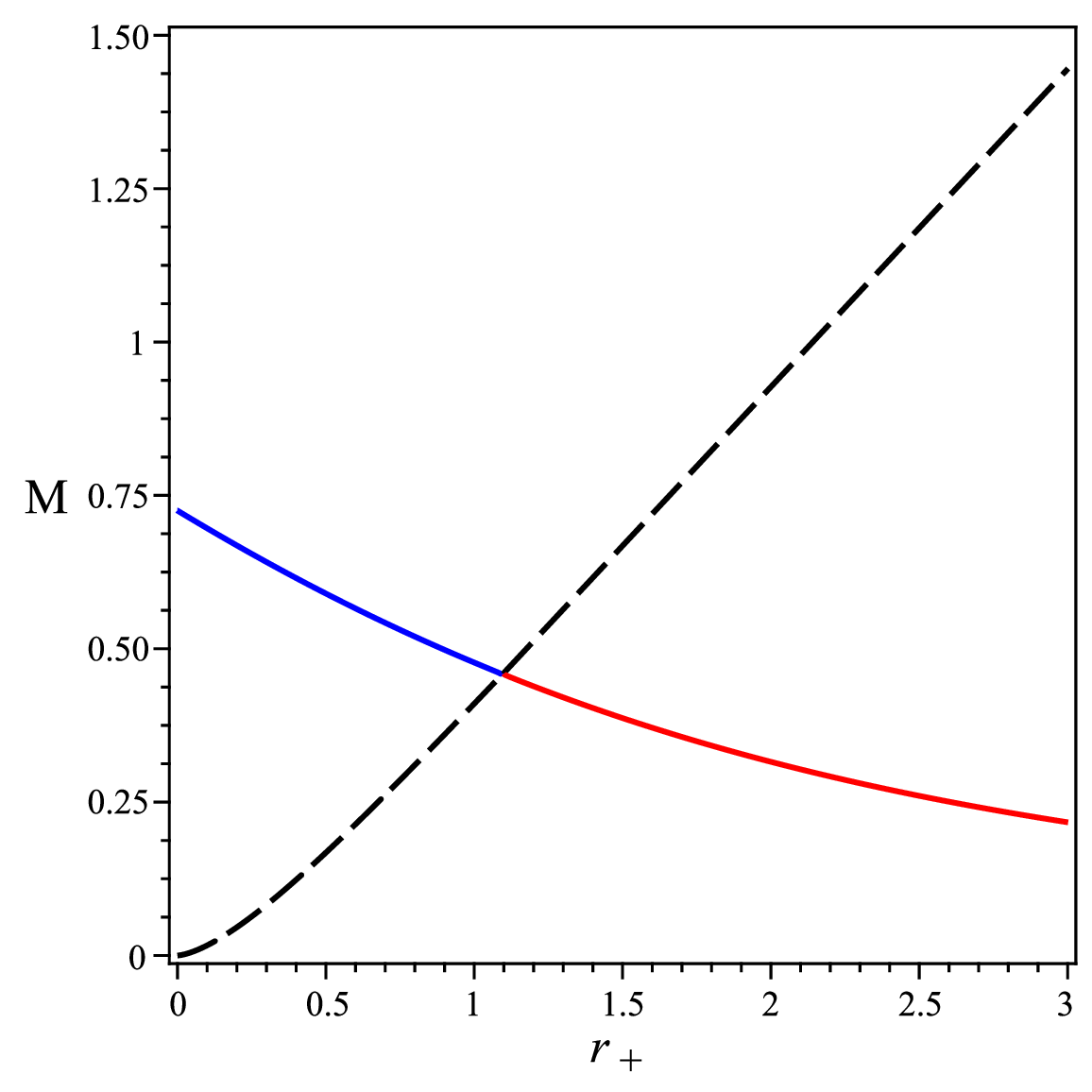}
 %\label{fignT01}
 }
 \subfigure%[$ c_{1}=1.052$]
 {
 \includegraphics[width=0.3\columnwidth]{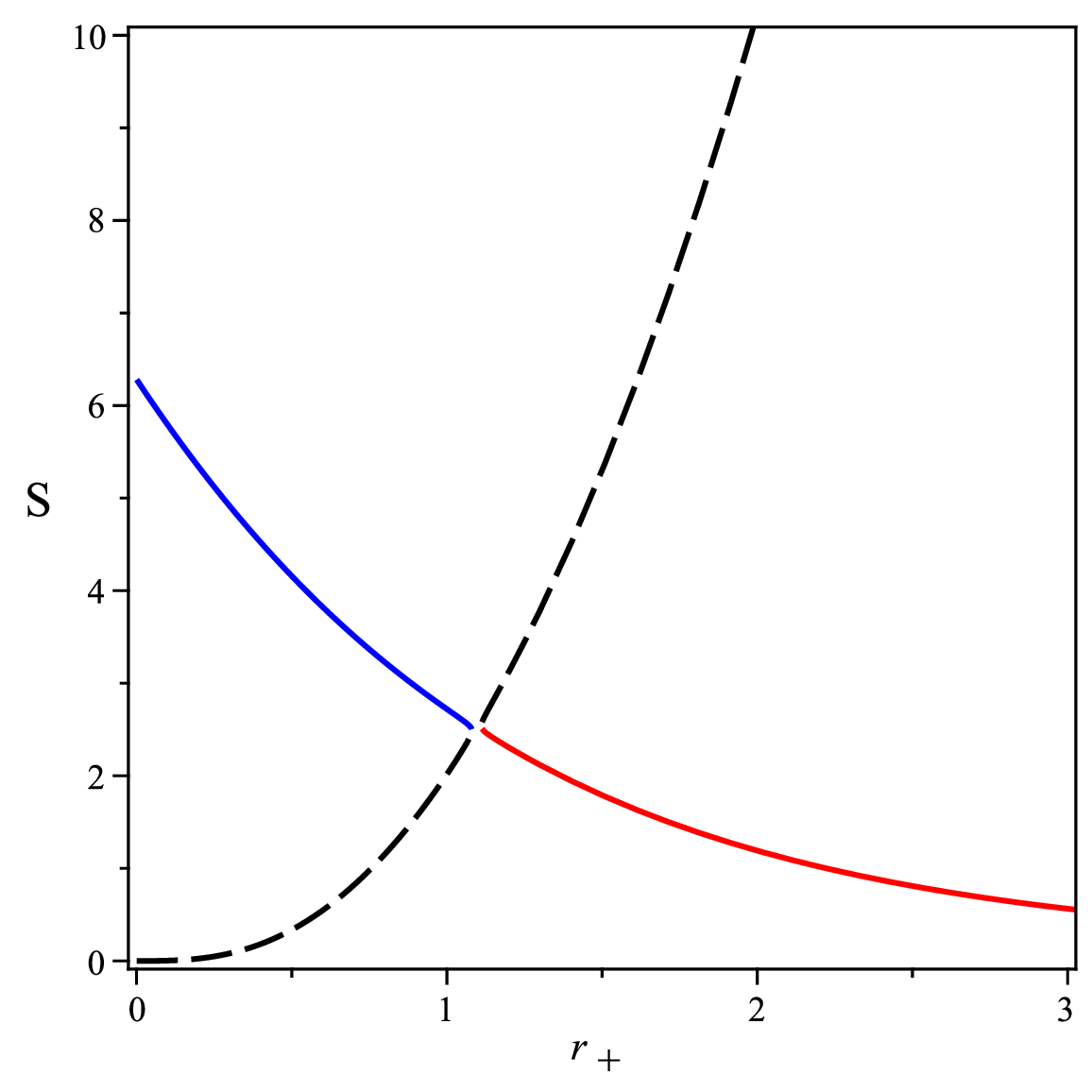}
 %\label{fignM1}
 }
 \subfigure%[$ c_{1}=1.052$]
 {
 \includegraphics[width=0.3\columnwidth]{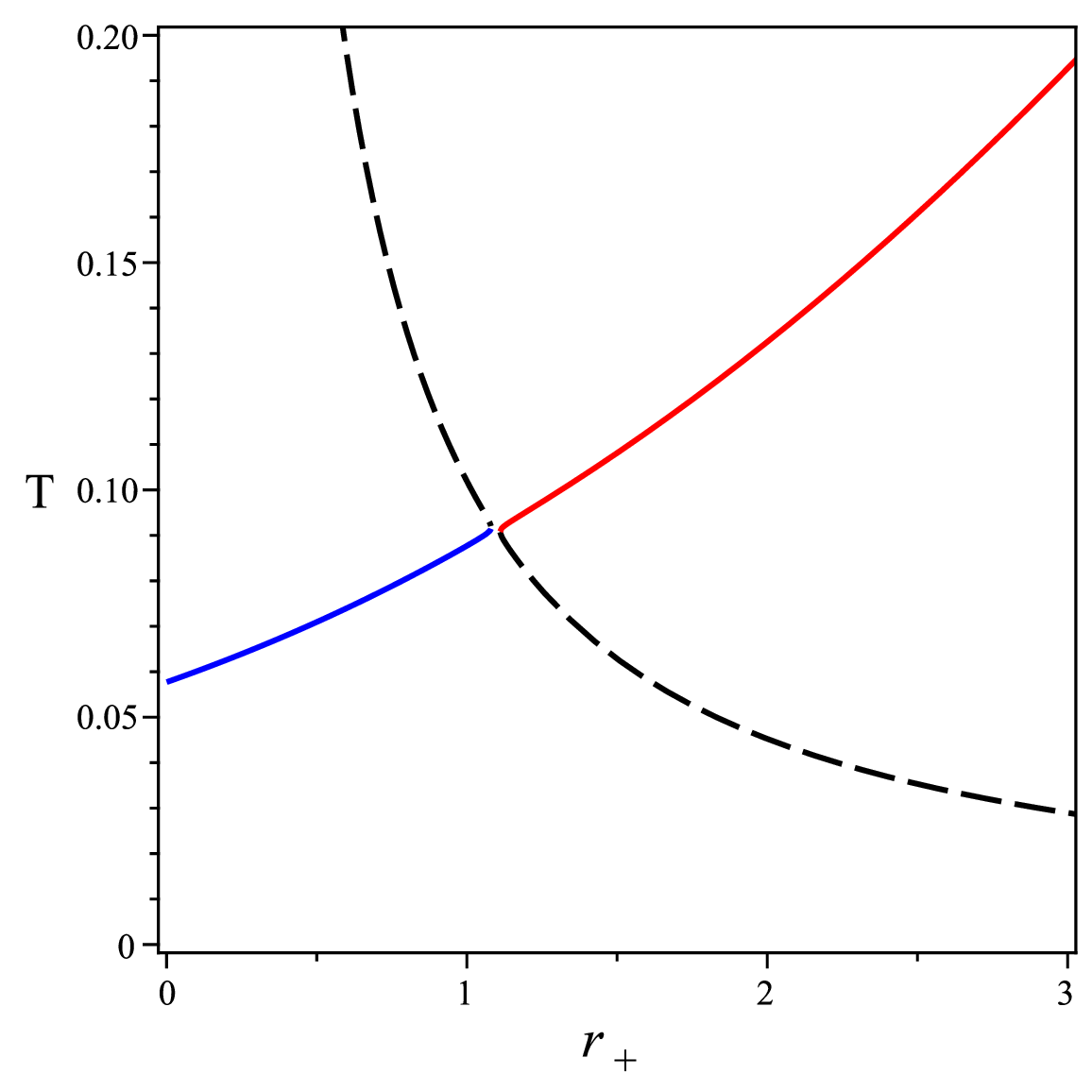}
 %\label{fignS1}
 }
 \caption{Plots of $ M $, $ S $ and $T$ in terms of $r_{+}$ for $\alpha=0.5, c_{1}=1.052$. 
The dashed line curves indicate Schwarzschild-like behavior and solid line curves are non-Schwarzschild-like behavior.}
\label{MSTrpplottasquarcrit}
 \end{figure}

\begin{figure}[h]\hspace{0.4cm}
\centering
\subfigure{\includegraphics[width=0.3\columnwidth]{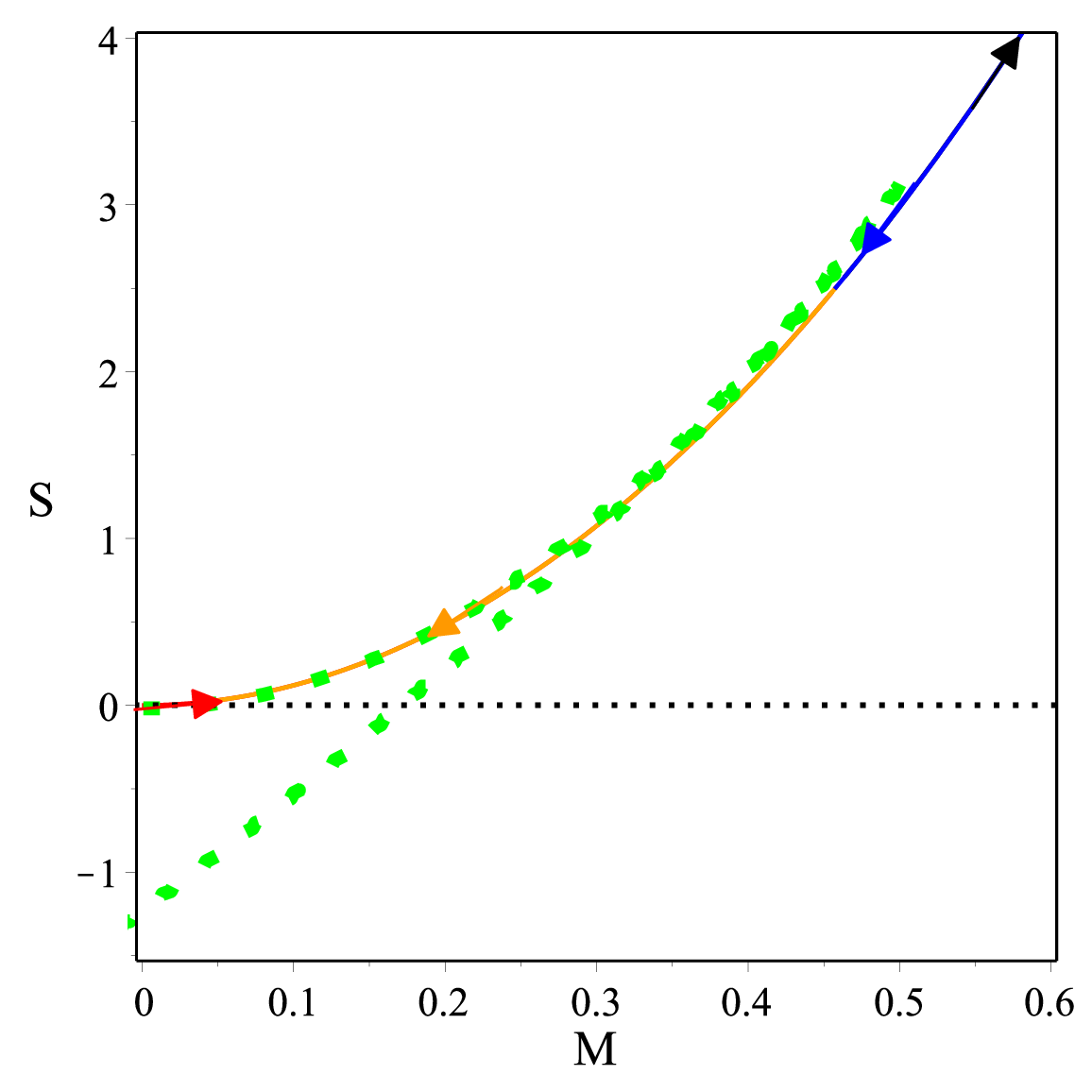}}
\subfigure{\includegraphics[width=0.3\columnwidth]{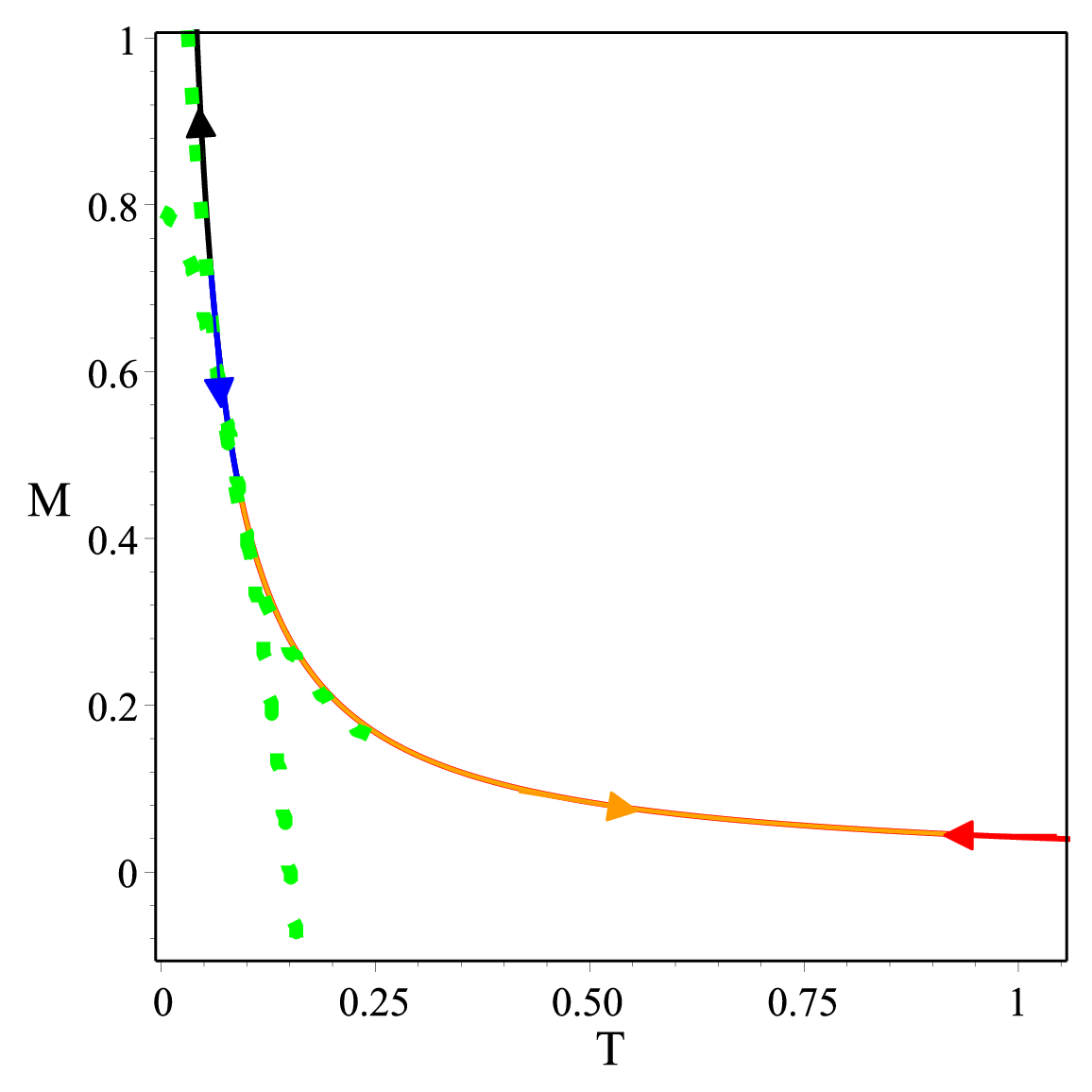}}
\subfigure{\includegraphics[width=0.3\columnwidth]{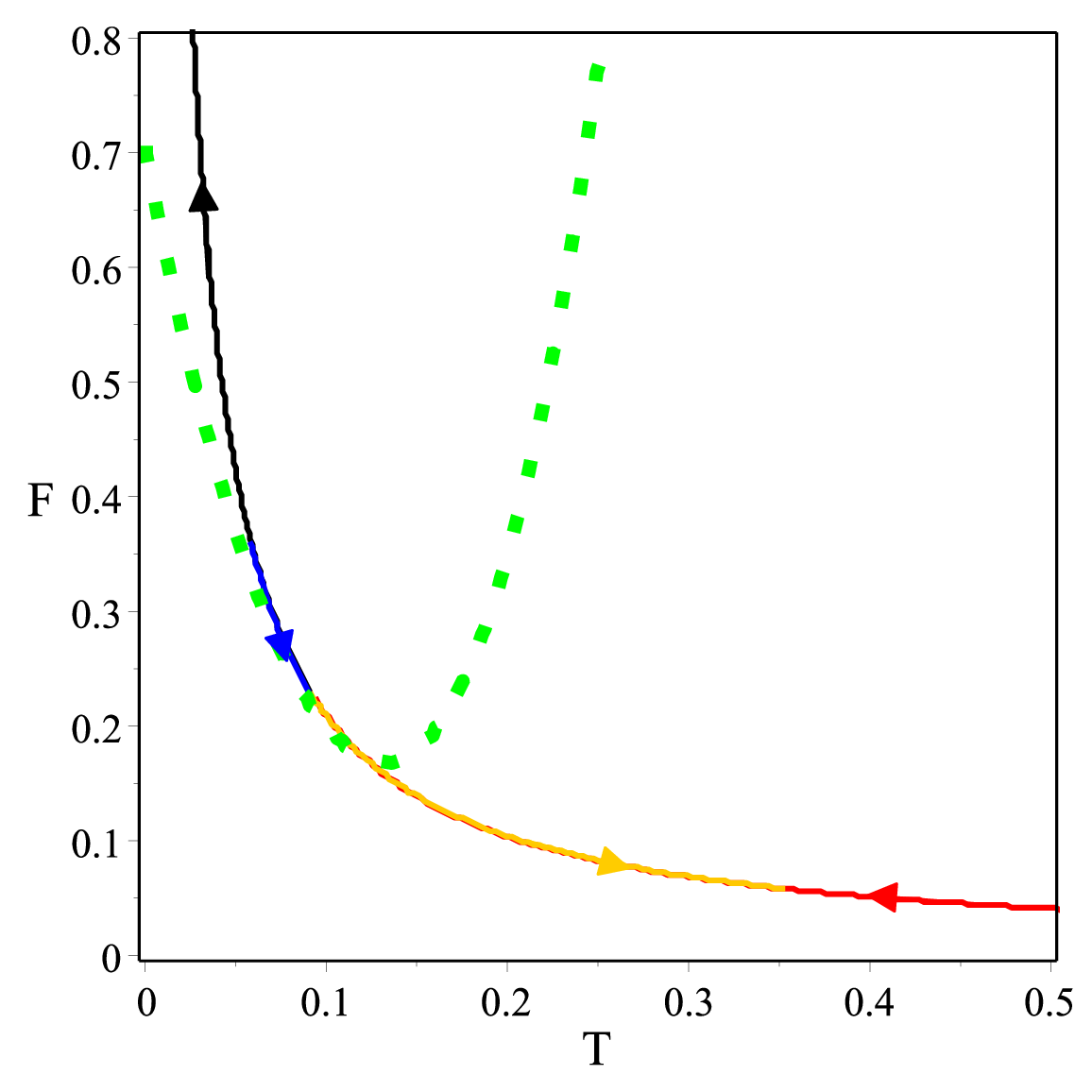}}
\caption{Left: Entropy $(S)$ as a function of black hole's mass $(M)$. Middle: Black hole's mass $(M)$ as a function of temperature $(T)$. Right: Free energy $(F)$ as a function of temperature $(T)$. For $ \alpha=0.5$ and  $c_1=1.052$. The direction of the arrows shows the direction of increasing $r_{+}$. The colors correspond to different branches of the solution. 
} 
\label{figmst1}
\end{figure}

\begin{figure}[H]
\centering
\subfigure%[]
{
 \includegraphics[width=0.3\columnwidth]{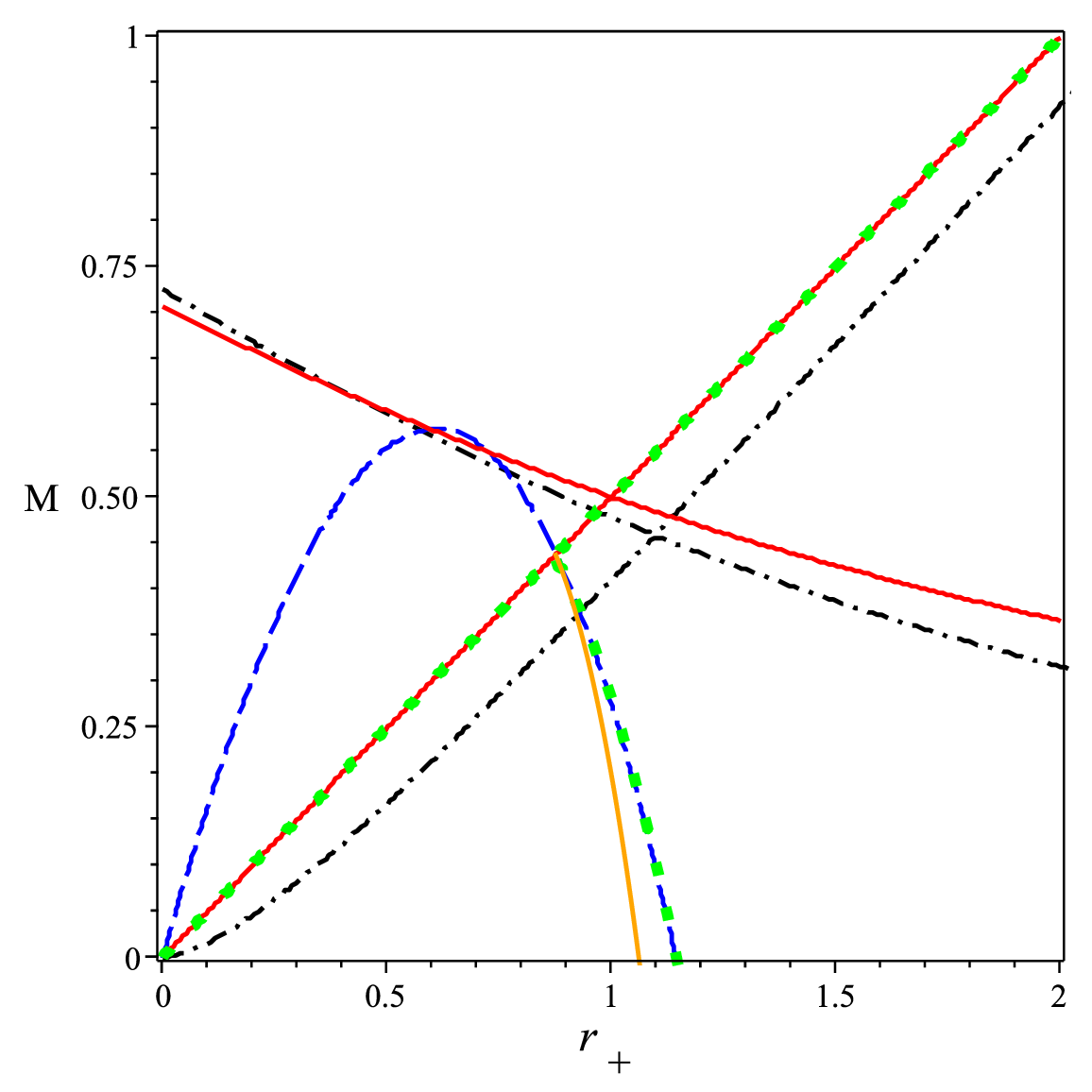}
 %\label{fignT0}
 }
 \subfigure%[]
 {
 \includegraphics[width=0.3\columnwidth]{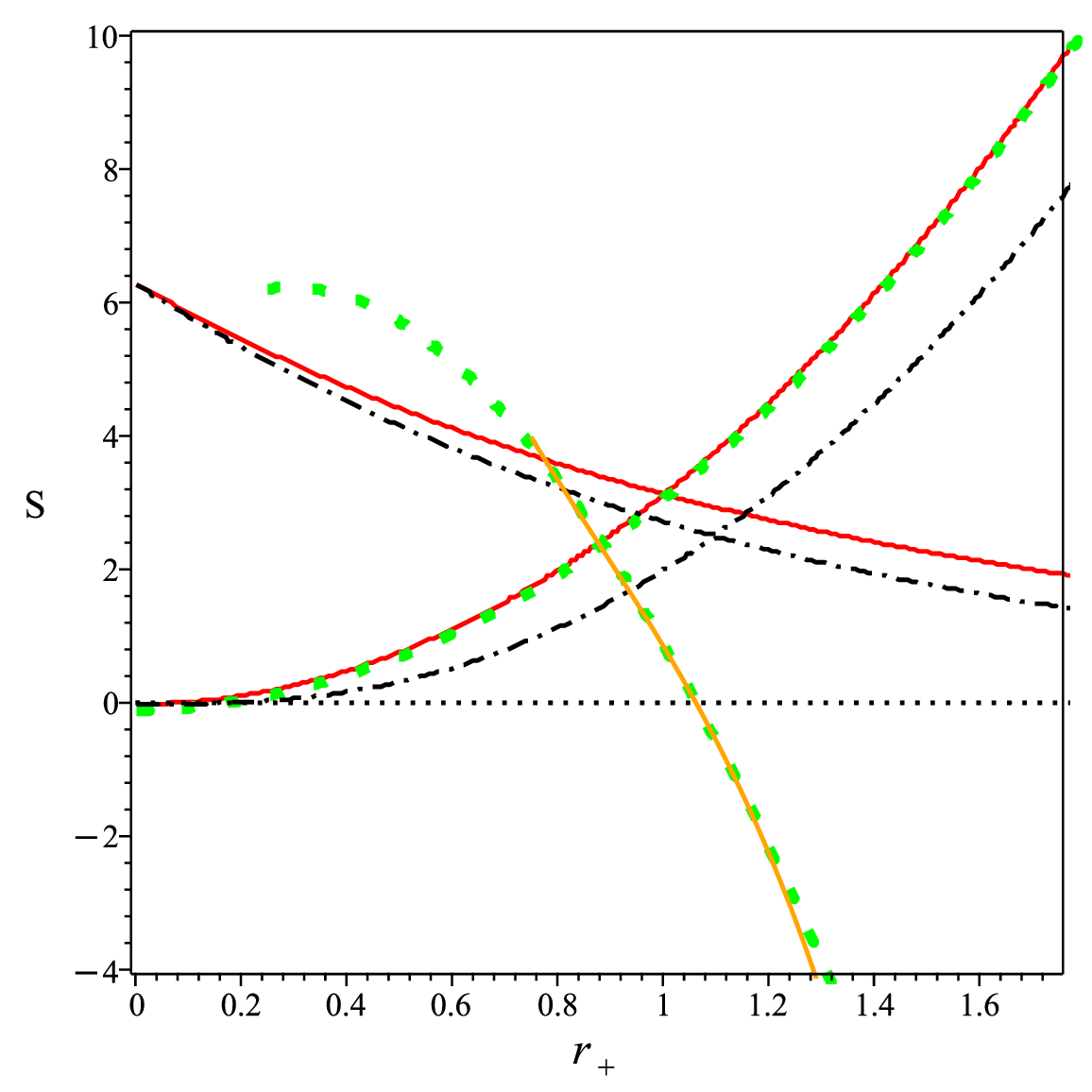}
 %\label{fignM0}
 }
 \subfigure%[]
 {
 \includegraphics[width=0.3\columnwidth]{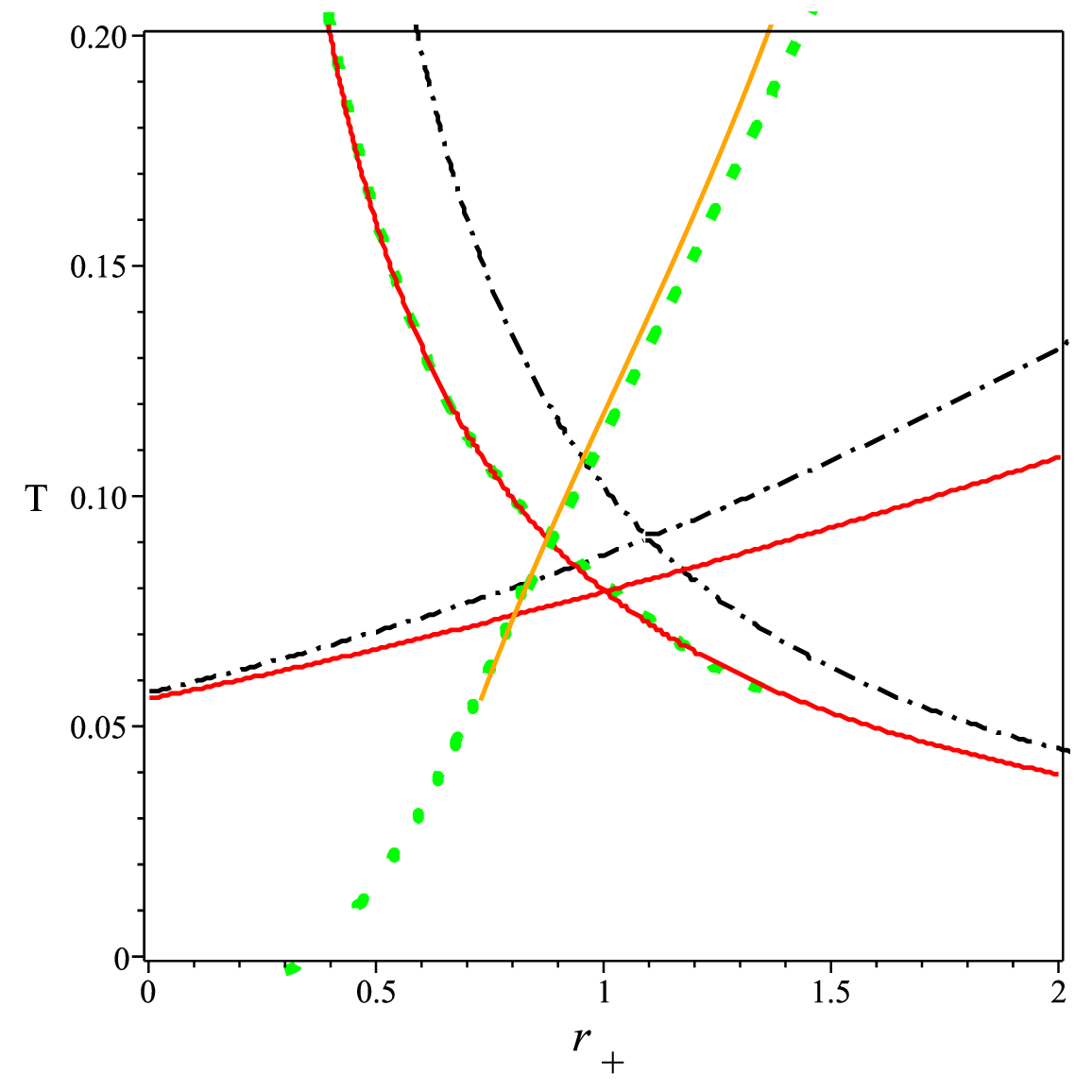}
 %\label{fignS0}
 }
 \caption{Plots of $ M $, $ S $ and $T$ in terms of $r_{+}$ for $\alpha=0.5$ and $c_1=1.052$. 
The black curves indicate Schwarzschild-like behavior and blue (cold) and red (hot) line curves denote non-Schwarzschild-like behavior. Solid and dashed lines correspond to the linear and square function between $h_{1}$ and $f_{1}$ respectively. Gold solid lines are thermodynamics quantities corresponding to \eqref{eqh131} and \eqref{eqf132}. }
\label{figmstcomb1}
\end{figure}

In figure \ref{CC0plot1} and the right panel of figure \ref{figmst1}, we have shown the behavior of heat capacity and free energy. {As can be seen, the heat capacity is negative for all values of $r_{+}$ which shows that black holes are locally unstable and this is in agreement with figure 6 of \cite{Lu:2017kzi}.} The free energy of the Schwarzschild-like solutions (red and black color) has increased in ${r_+}$, while non-Schwarzschild-like solutions (blue and green color) have decreased in ${r_+}$. This shows that the non-Schwarzschild-like solutions globally are stable. 

Inserting \eqref{eqq2324}-\eqref{eqq24} into \eqref{eqq11}, one can obtain an analytic solution for the metric of EQG which is valid everywhere outside the horizon. 
We have shown the metric functions $f(r),h(r)$ in figure \ref{FHplot0}. These are asymptotically flat non-Schwarzschild-like black holes.
It should be noted that there is no peak in figure \ref{FHplot0}. Because, as one can see in the figure \ref{MSTrpplottasquar1} the mass of non-Schwarzschild-like black holes for squared relation between $h_{1}$ and $f_{1}$ is always positive \cite{Lu:2015cqa}. In the right panel of figure \ref{FHplot0}, we have shown the difference between numerical and analytical approximation for the metric functions. Similar to the linear case ($f_1=h_1$), in the small and large radius, there is a good agreement between numerical and analytical results. There is about $10\%$ difference around $r\approx 3.5$.

\begin{figure}[h]
\centering
\subfigure{\includegraphics[width=0.4\columnwidth]{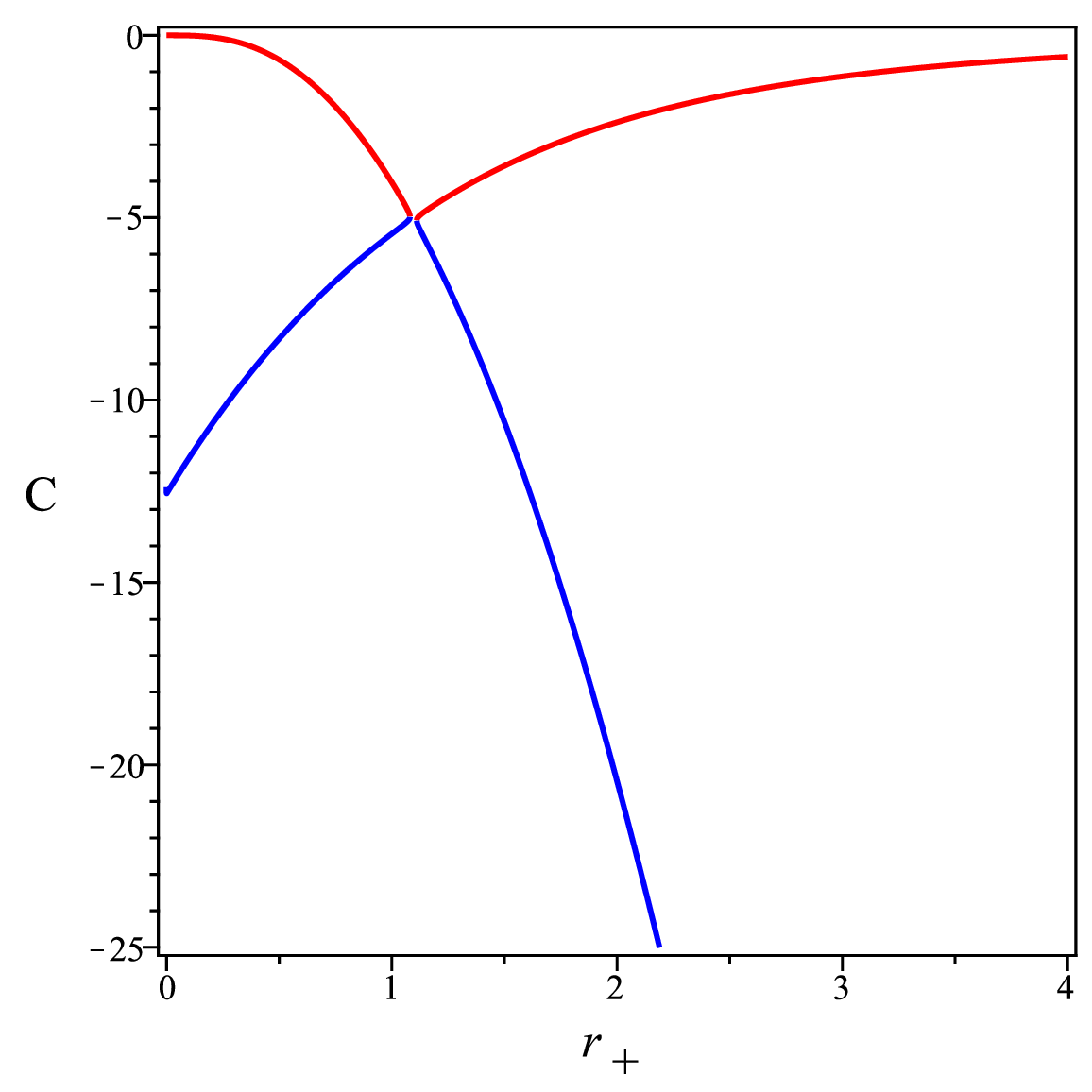}}
%\subfigure[$c_{1}=1.052$]{\includegraphics[width=0.4\columnwidth]{FTTplot1}}
\caption{\small
The behavior of heat capacity in terms of $r_{+}$, for $ \alpha=0.5$ and  $c_1=1.052$.}
\label{CC0plot1}
\end{figure}

\begin{figure}[h]
\centering
\subfigure[$\alpha=0.5, c_{1}=1.052, r_{+}=1$]{\includegraphics[width=0.49\columnwidth]{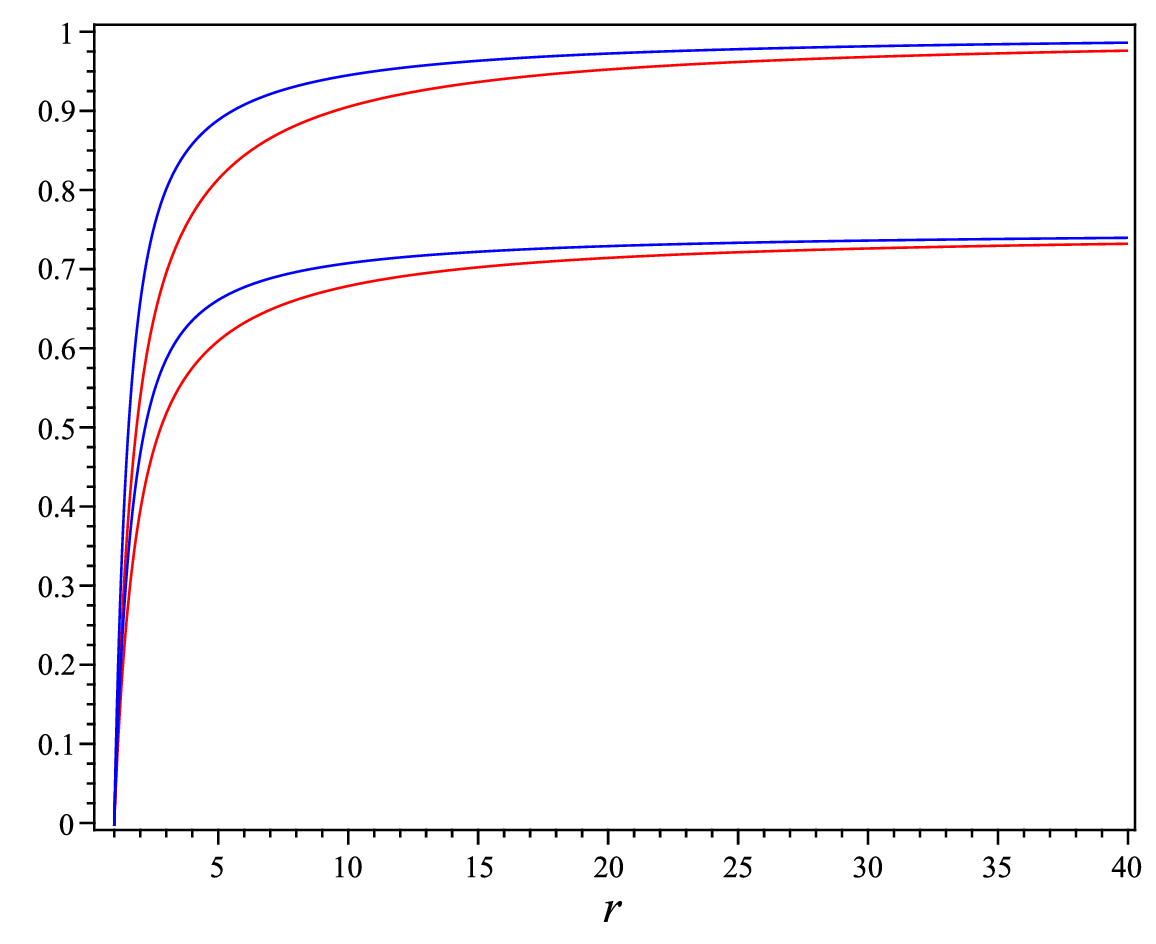}}
\subfigure[$\alpha=0.5, c_{1}=1.052, r_{+}=1$]{\includegraphics[width=0.4\columnwidth]{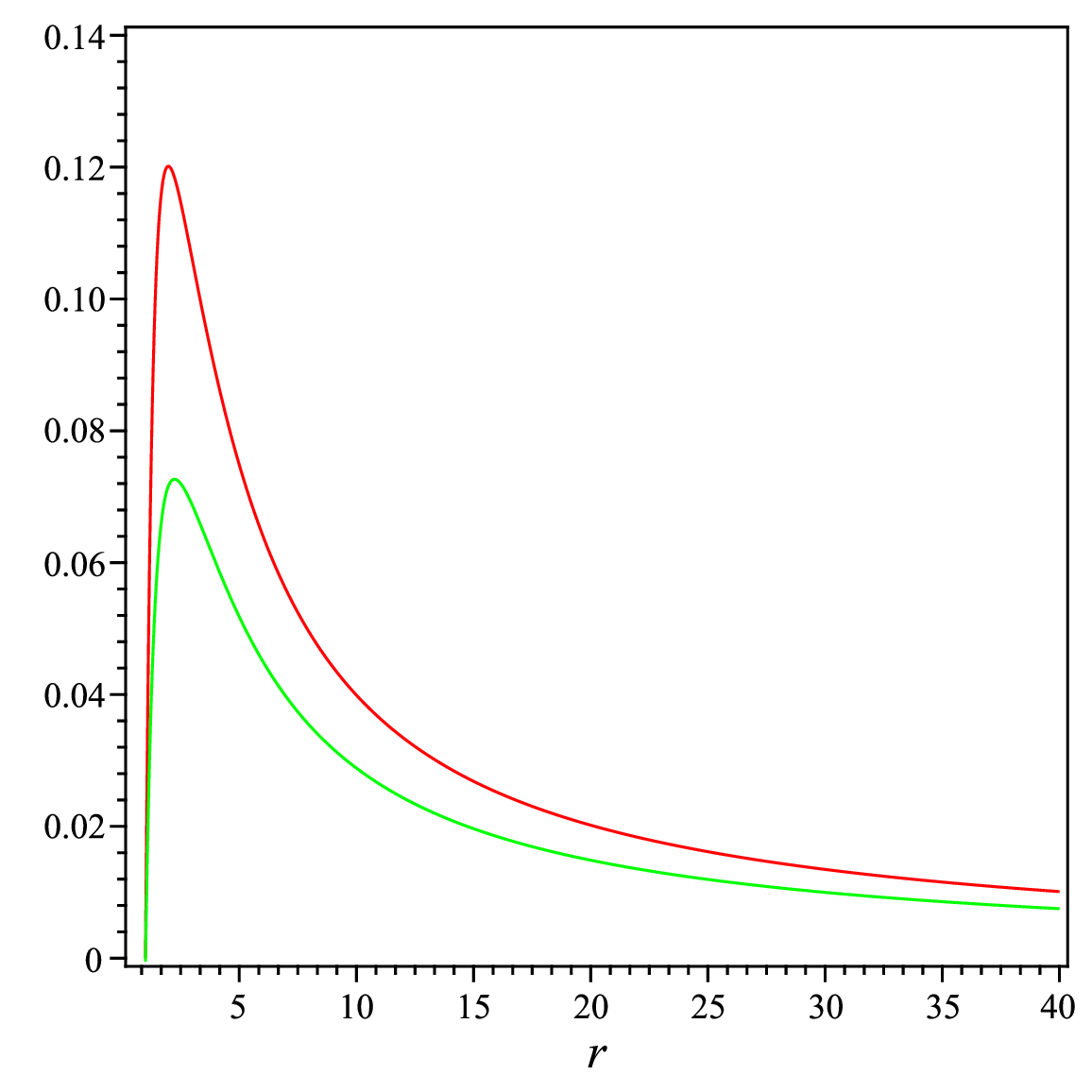}}
%\subfigure[$r_{+}=3, M=0.217$]{\includegraphics[width=0.3\columnwidth]{fhplott}}
\caption{Left: The behavior of $h(r)$ and $0.75f(r)$ (\textcolor{blue}{blue lines},upper) of our approximated solutions. The numerical $h(r)$ and $0.75f(r)$ (\textcolor{red}{red lines}) according to \cite{Kokkotas:2017zwt}(lower). Right: The difference between analytical and numerical approximations for $h(r)$ (\textcolor{red}{red line}, upper) and $f(r)$ (\textcolor{green}{green line}, lower).} 
\label{FHplot0}
\end{figure}

In this section, we examine the case of $f_{1}\neq h_{1}$ and find that the thermodynamic quantities of our solutions deviate from the Schwarzschild behavior. We discover that if the proper behavior of $h_{1}$ and $f_{1}$ are known, the numerical results related to the non-Schwarzschild black hole solutions can be obtained. {However, we still believe that non-Schwarzschild branches related to $h_1=f_{1}$ and $h_{1}=f_{1}^2$ are non-Schwarzschild black holes solutions. Although they are different from the numerical calculations done so far.}\\

In the next subsection, we examine the dynamical stability of novel black holes related to different branches (small-non-Schwarzschild and large non-Schwarzschild) of the case $h_{1}=f_{1}^2$.

\subsection{Dynamical Stability}\label{sub22}

Here, we consider a massless scalar field $\Phi$ propagating in a black hole background governed by the Einstein-Weyl theory. Its evolution obeys the massless Klein-Gordon equation
\begin{equation}
\square \Phi=0.
\end{equation}

To solve the above Klein-Gordon equation analytically, it is convenient to use the
following the standard separation of variables by making use of the spherical harmonic functions $(Y_{lm})$
\begin{equation}
\Phi(t,r,\theta, \phi)=\sum_{lm}\dfrac{1}{r}\varphi_{l}(r)Y_{lm}(\theta,\phi)e^{i\omega t}.
\end{equation}
Accordingly, the Klein-Gordon equation can be simplified as
\begin{equation}
r^2 f\partial^{2}_{r}\varphi_{l} + \left(\dfrac{r^{2}\partial_{r}f}{2}+\dfrac{r^2f\partial_{r}h}{2h}\right)\partial_{r}\varphi_{l} +  \left(\dfrac{r^2 \omega^2}{h}-\dfrac{r\partial_{r}f}{2}-\dfrac{rf\partial_{r}h}{2h}-l(l+1)\right)\varphi_{l} =0,
% +
% \left(\dfrac{r^{2}\partial_{r}f}{2}+\dfrac{r^2f\partial_{r}h}{2h}\right)\partial_{r}\varphi_{l}+r^2 f\partial^{2}_{r}\varphi_{l}=0,
\end{equation}
where $l$ is the spherical harmonic index. By introducing the tortoise coordinate,
\begin{equation}
dr^{\ast}=\dfrac{dr}{\sqrt{f(r)h(r)}},
\end{equation}
we obtain the Schr\"odinger-type equation as follows,
\begin{equation}\label{shreq}
\left(\partial^{2}_{r^{\ast}}+\omega^2 -U_{l}(r)\right)\varphi_{l}(r^{\ast})=0,
\end{equation}
where the effective potential is given by
\begin{equation}
U_{l}(r(r^{\ast}))=\dfrac{1}{2r}\partial_{r}\left(fh\right)+\dfrac{l(l+1)h}{r^2}. \label{Ueff}
\end{equation}
\begin{figure}[H]\hspace{0.4cm}
\centering
\subfigure[$c_{1}=1.052,r_{+}=1,\alpha=0.5$]{\includegraphics[width=0.45\columnwidth]{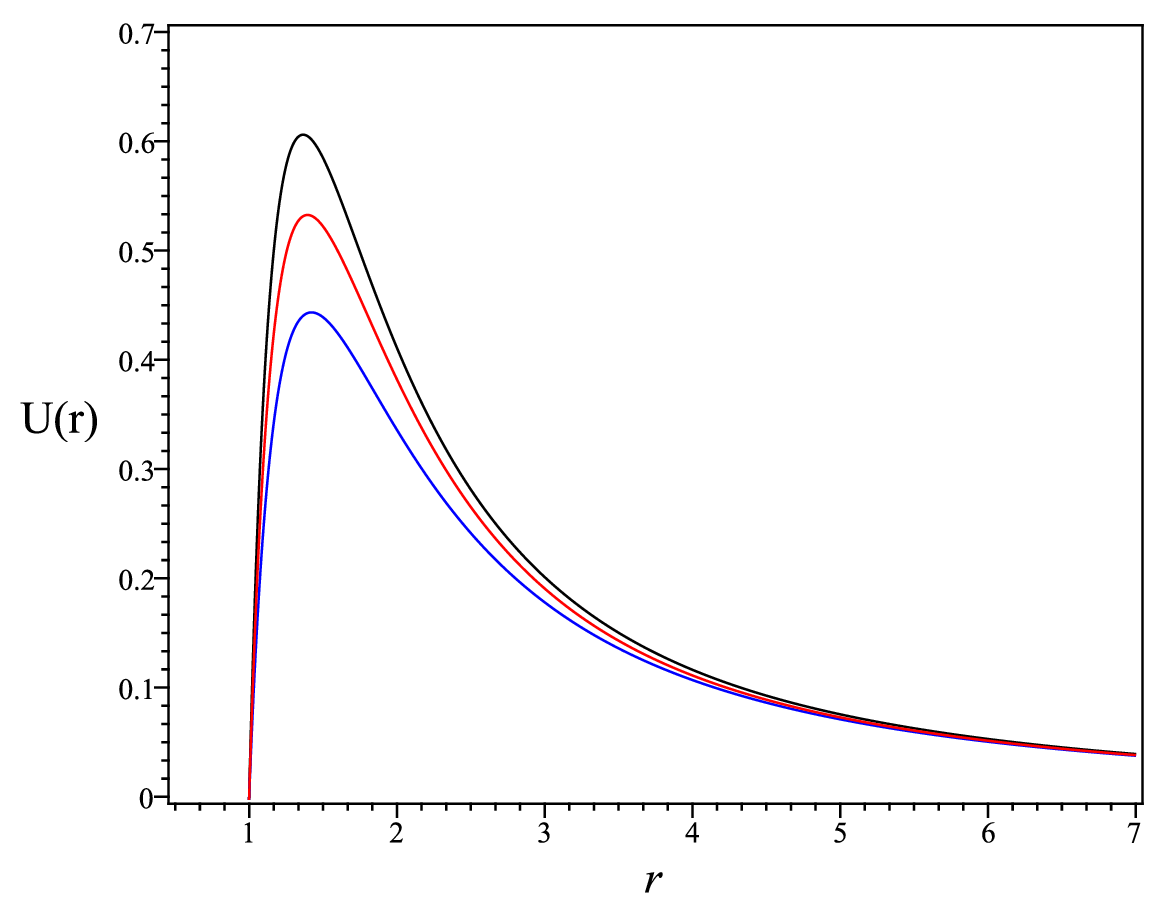}}
\subfigure[$c_{1}=1.052,r_{+}=1,\alpha=0.5$]{\includegraphics[width=0.45\columnwidth]{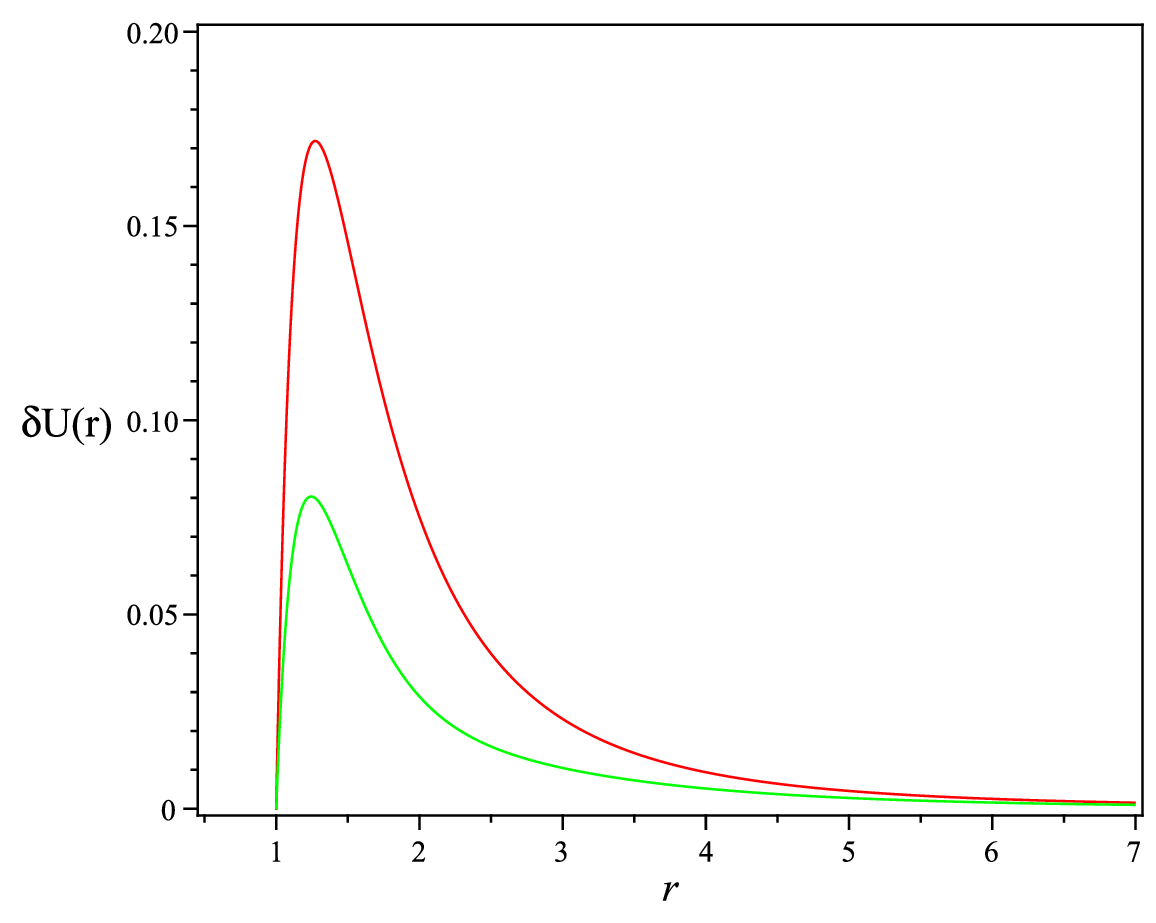}}
\caption{\small
Left: The behavior of effective potential $U(r)$ in terms of $r$ for \textcolor{black}{black line} (upper, numerical solution), \textcolor{red}{red line} (middle, Schwarzschild-like) and \textcolor{blue}{blue line} (lower, non-Schwarzschild-like). Right: The difference between analytical and numerical approximations of $\delta U(r)$ (\textcolor{red}{red line}, upper) (Schwarzschild-like) and $\delta U(r)$ (\textcolor{green}{green line}, lower) (non-Schwarzschild-like).}
\label{UUr0plot1}
\end{figure}

In figure \ref{UUr0plot1}, the effective potential for analytical and numerical black holes \cite{Kokkotas:2017zwt} are plotted and compared. It can be seen from the right panel that the discrepancy for non-Schwarzschild black holes is less than 10$\%$ around $r\approx 1.3$. In the eikonal limit (large $l$), the effective potential becomes \cite{Kokkotas:2017ymc}
\begin{equation}
U(r)\approx \left(l+\dfrac{1}{2}\right)^2\dfrac{h(r)}{r^2},
\end{equation}
with 
\begin{equation}\label{eqhapprox}
h(r)\approx \left(1-\dfrac{r_+}{r}\right)\left[1+\dfrac{3r_{+}^3(41r+11r_{+})p^2}{r^2(215r^2-340r_{+}r+437r_{+}^2)}\right]+\mathcal{O}(p^3).
\end{equation}
We denote $r_p$ as the radius at which the potential reaches its maximum i.e., $U'(r_p)=0$. Therefore, in the eikonal limit, $r_p$ can be obtained 
\begin{equation}\label{eqrpapprox}
r_{p}=\dfrac{3r_{+}}{2}-\dfrac{1079462r_{+}p^2}{8098347}+\mathcal{O}(p^4).
\end{equation}
The peak of the effective potential is closer to the black hole horizon for non-Schwarzschild black holes than for the Schwarzschild one. 
%The behavior of $r_{p}$ in terms of $2M$ is shown in figure \ref{rpsplot}. {On the left panel, we compare our approximate black hole solutions (red and blue solid lines) with numerical black hole solutions (dotted green curve \cite{Pravda:2024uyv}  and black dashed curve \cite{Kokkotas:2017zwt}). We observe that our solution and numerical solution are qualitatively similar for the Schwarzschild-like branch. The discrepancy becomes more visible at large $M$. In contrast, our solution does not agree with the numerical one for the non-Schwarzschild-like branch. In our case, we find no evidence of a negative black hole's mass.}

%\begin{figure}[H]\hspace{0.4cm}
% \centering
 %\subfigure{\includegraphics[width=0.45\columnwidth]{rpsmplot11}}
 %\subfigure{\includegraphics[width=0.45\columnwidth]{wrwiplott}}
 %\caption{The behavior of $r_{p}$ in terms of $2M$ for $c_{1}=1.052,\alpha=0.5$. The red curve denotes Schwarzschild-like behavior and the blue curve denotes non-Schwarzschild-like behavior. The green dotted lines and dashed lines correspond to numerical solutions in \cite{Pravda:2024uyv} and \cite{Kokkotas:2017zwt}, respectively (left). The behavior of real part $\omega_{r}$ (solid lines) and imaginary part $\omega_{i}$ (dashed lines) of QNM in the eikonal limit (right).} 
% \label{rpsplot}
% \end{figure}
\noindent We solve equations \eqref{shreq} with \eqref{Ueff} and investigate the dynamical stability of the non-Schwarzschild black hole solutions by exploring the quasinormal mode of scalar perturbation. The corresponding quasinormal frequencies are obtained via the Method provided in \cite{vf, Mashhoon:1985cya}. In tables \eqref{table2}--\eqref{table23}, quasinormal modes of small and large non-Schwarzschild-like black holes are listed. First, from the tables \eqref{table2} and \eqref{table22}, one can conclude that in the valid interval of $r_{+}$, all quasi-normal frequencies have a positive imaginary part and large/small non-Schwarzschild black holes are stable. It should be noted that for the small non-Schwarzschild black hole from $r_{+}=0.6$ to $r_{+}=1.0$, we find that the quasi-normal frequency does not have a real part and only has a negative imaginary part. Therefore, the black holes with the event horizon in the mentioned range are unstable ($0.6<r_{+}<1.0$).  It is also observed from the tables \eqref{table1} and \eqref{table23} for small and large black holes, with increasing $l$, the values of the real part of the frequencies increase while the imaginary part of the frequencies decreases (the same as Schwarzschild black holes). Moreover, we also find that $r_p$ increases with $l$ as expected. Figure \ref{mrplotstable} shows a summary of the black holes' dynamical stability and instability ranges. The blue lines and red lines are stable and black lines are unstable.

\begin{table}[H]
\begin{center}
\setlength{\tabcolsep}{0.7cm}
\begin{tabular} { c  c  c  c  c}
\hline
 $r_{+}$&$\omega_{r}$&$\omega_{i}$ \\
 \hline
 2&0.5226042685 &0.327440433\\
 
 2.5&0.4841843547 &0.4565269333 \\
  
 3.0 &0.4642678481&0.575397849 \\
   
3.5&0.4432288172&0.7040145415 \\

4.0 &0.3668954701&0.8676194445 \\

4.5&0.3622784854&0.9987396000 \\

%5.0&0.03101977032&0.03840195759 \\

\hline
\end{tabular}
\end{center}
\caption{ The values for $r_{+}$ and the quasinormal frequencies in the (large) non-Schwarzschild geometry (red branch of a figure (\ref{MSTrpplottasquarcrit})) for the values of parameters $\alpha=0.5, c_{1}=1.052,l=1,n=0$. \label{table2}}
\end{table}

\begin{table}[H]
\begin{center}
\setlength{\tabcolsep}{0.7cm}
\begin{tabular} { c  c  c c c}
\hline
 $l$&$r_{p}$&$\omega_{r}$ &$\omega_{i}$ \\
 \hline
% 0&9.849999979&0.01078231800&0.1662364092&0.1298161228 \\
 
1&3.607963697&0.4642678481&0.575397849 \\
  
 2&3.682050904 &0.8162055214 &0.5371547855 \\
   
3&3.756683235&1.151557667&0.5069365175 \\

4&3.803098300 &1.479587505 &0.4910348779 \\

5&3.833145397&1.806073768&0.4818340994 \\

6&3.851130728&2.132053936&0.4768797332 \\

7&3.864060551&2.458143688&0.4734296248 \\

8&3.873007350&2.784302481&0.471143328 \\
\hline
\end{tabular}
\end{center}
\caption{ The values for $r_{p}$ and the quasinormal frequencies in the (large) non-Schwarzschild geometry for the values of parameters $r_{+}=3,\alpha=0.5, c_{1}=1.052,n=0$. \label{table1}}
\end{table}

\begin{table}[H]
\begin{center}
\setlength{\tabcolsep}{0.7cm}
\begin{tabular} { c  c  c  c  c}
\hline
 $r_{+}$&$\omega_{r}$&$\omega_{i}$ \\
 \hline
 %0.05&0.8342924415&0.2851283928\\
 
 0.1&0.4227173939&0.1465573637\\
 
 0.2&0.4210530639 &0.1608539126 \\
  
 0.3 &0.4085568132&0.1830207403 \\
   
0.4&0.3887830967&0.2178240416 \\

0.5&0.3518733013&0.284973687 \\

0.6&0.200698489&0.434552487 \\

%0.50&0.7168966761&0.7950611755 \\

%0.53&0.1013832866&0.9239348375\\
\hline
\end{tabular}
\end{center}
\caption{ The values for $r_{+}$ and the quasinormal frequencies in the (small) non-Schwarzschild geometry for the values of parameters $\alpha=0.5, c_{1}=1.052,l=1,n=0$. From $r_{+}=0.6$ to $r_{+}=1.0$, the small non-Schwarzschild black hole are unstable.\label{table22}}
\end{table}

\begin{table}[H]
\begin{center}
\setlength{\tabcolsep}{0.7cm}
\begin{tabular} { c  c  c  c  c}
\hline
 $l$&$r_{p}$&$\omega_{r}$ &$\omega_{i}$ \\
 \hline
  
1& 2.156471888&0.4158489387&0.1706211364 \\
  
 2&2.234360749&0.6793968775&0.1638160176 \\
   
3&2.261788601&0.9455008131&0.1621428624 \\

4&2.273922288 &1.212657043&0.1614951428 \\

5&2.280257633&1.480297174&0.1611772776 \\

6&2.283960874&1.748196373&0.1609976096 \\

7&2.286306599&2.016250026&0.1608860806 \\

8&2.287883974&2.284403121&0.1608120464 \\
\hline
\end{tabular}
\end{center}
\caption{The values for $r_{p}$ and the quasinormal frequencies in the (small) non-Schwarzschild geometry (blue branch of the figure (\ref{MSTrpplottasquarcrit}))  for the values of parameters $r_{+}=0.25,\alpha=0.5, c_{1}=1.052,n=0$. \label{table23}}
\end{table}

\begin{figure}[H]
\centering
\subfigure{\includegraphics[width=0.5\columnwidth]{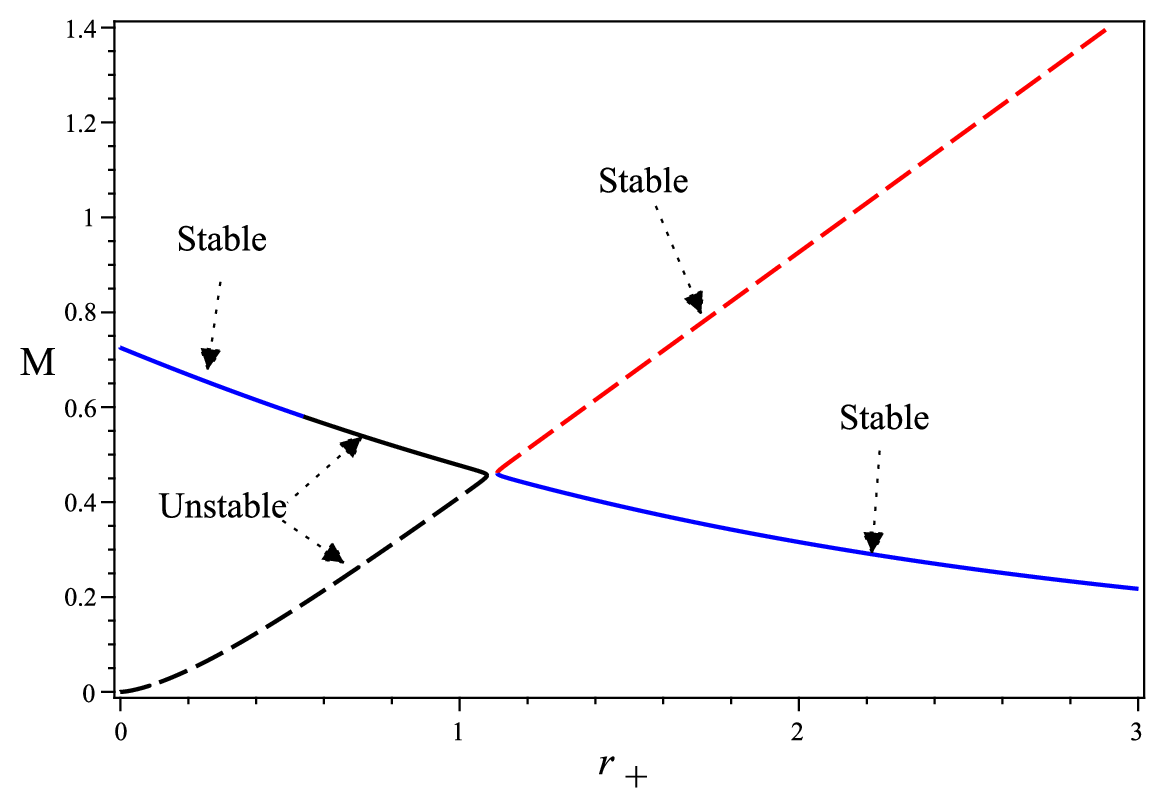}}
%\subfigure[$c_{1}=1.052$]{\includegraphics[width=0.4\columnwidth]{FTTplot1}}
\caption{\small
Dynamical stability ranges for Schwartschild (dashed line) and non-Schwartschild (solid line) black holes.}
\label{mrplotstable}
\end{figure}
 
In the eikonal limit, the quasinormal frequencies can be approximately given as \cite{Kokkotas:2017zwt}
\begin{align}
% &\;\;\;\;\;\;\;\;\;\;\;\;\;\;\;\;\;\;\;\;\;\;\;\;\;\;\
\omega&=\sqrt{U(r_{p})}+i\left(n+\dfrac{1}{2}\right)\left.\sqrt{-\dfrac{1}{2U}\dfrac{d^2U}{dr_{\ast}^2}}\right\vert_{r_{p}}\nonumber\\
&=\left(l+\dfrac{1}{2}\right)\left[\dfrac{2\sqrt{3}}{9r_{+}}+\dfrac{1160\sqrt{3}p^2}{44361r_{+}}\right]+i\left(n+\dfrac{1}{2}\right)\left[\dfrac{2\sqrt{3}}{9r_{+}}+\dfrac{24681262640\sqrt{3}p^2}{359250771267r_{+}}\right]+\mathcal{O}(p^4).
\end{align}

%{We plot the quasinormal frequencies as a function of ${r_+}$ in the right panel of figure \ref{rpsplot}.} The behavior of real and imaginary parts of QNM in the eikonal limit has been shown. In this panel, the red lines are quasinormal frequencies of massless scalar field perturbing our analytic black hole while the blue lines represent those of the numerical one \cite{Kokkotas:2017zwt}. Solid lines denote the real part of the frequency while dashed lines illustrate the imaginary part. Clearly, the plots show dissimilarity of dynamical stability between our analytical solutions and the numerical ones \cite{Kokkotas:2017zwt}. For our black hole solutions, both frequency parts decrease with ${r_+}$. On the other hand, in the numerical solution \cite{Kokkotas:2017zwt}, both frequency increases with ${r_+}$. This is justified because here we considered a simple relationship between $f_{1}$ and $h_{1}$ compared to \eqref{eqh131} and \eqref{eqf132}.

Usually, the eikonal QNM of a test scalar field determines the parameters of the circular null orbits around the black hole. The real part of the QNM has a relationship with the shadow radius. In the following, we study the shadow cast by the photon sphere of the black holes. An unstable circular orbits of null geodesics $r_{ps}$ and shadow radius $R_{sh}$ are given as \cite{Mafuz:2023qxn}
\begin{eqnarray}
   \left.\dfrac{h^{\prime}(r)r}{h(r)}\right\vert_{r=r_{ps}}=2,\;\;\;\;\;\;\;R_{sh}^2=\left.\dfrac{r^2}{h(r)}\right\vert_{r=r_{ps}}, \label{ShadowRadius}
\end{eqnarray}
At large $l$, it is easy to show that $r_{ps}=r_{p}$. Therefore, by using \eqref{eqhapprox} and \eqref{eqrpapprox} we achieve  
\begin{equation}
R_{sh}=\dfrac{3\sqrt{3}r_{+}}{2}-\dfrac{290\sqrt{3}r_{+}p^2}{1643}+\mathcal{O}(p^4).
\end{equation}
It is easy to show that the real part of QNM in the large $l$ limit is inversely proportional to the shadow radius $w_{r}= l/R_{sh}$. This connection is a reflection of the fact that the gravitational waves are treated as massless particles propagating along the last null unstable \cite{Jusufi:2019ltj,Jusufi:2020dhz,Hendi:2020knv}.

\section{Conclusion}\label{conc}

Following the paper \cite{Lu:2015cqa, Bonanno:2019rsq, Sajadi:2020axg}, we have obtained an analytic approximation for black hole solutions in Einstein quadratic gravity by making use of a continued-fraction expansion for a positive generic coupling constant $\alpha$. We consider the cases where $h_{1}=f_{1}$ and $h_{1}=f_{1}^{2}$. In the absence of a cosmological constant, we have obtained an analytic solution for an unknown function of the near horizon using the first law of thermodynamics and the Smarr formula in its proper form. By inserting this function into the metric functions and other thermodynamic quantities, we obtained analytical solutions for each case. In each case, we find at least four branches of the black hole solution including a Schwarzschild-like and non-Schwarzschild-like black hole. It should be noted that our results for the thermodynamics of the black hole are not perturbative and they have been obtained for a generic coupling constant.  
In addition to the coupling constant of the theory, there is an additional integration constant. The existence of black hole solutions depends on this constant integration. For $h_{1}=f_{1}$ and $h_{1}=f_{1}^{2}$, we have four branches of solutions, and for the cubic and other polynomial relations, extra nonphysical branches appear. For the linear case, i.e. $h_{1}=f_{1}$, our analytical results agree with the numerical results for the Schwarzschild-like branch. In \eqref{eqh131} and \eqref{eqf132}, we explicitly find the expression of $h_1$ and $f_1$. This gives us almost the numerical results of the non-Schwarzschild branch. To obtain numerical results with this analytical method, it is necessary to precisely know the relation between $h_1$ and $f_1$.

We also study the dynamical stability of the solutions by exploring quasi-normal modes. We find that the large non-Schwarzschild-like black hole solutions are stable dynamically. We found that the small non-Schwarzschild-like black hole solutions are not unstable for all values of the event horizon radii. It has a stable part in a small event horizon radius}. However, both black holes are unstable thermodynamically.

We leave for the future, obtaining the non-asymptotic flat black hole solution of the theory using this method and continued fraction expansion with different parameterizations \cite{Konoplya:2022kld}. Moreover, investigating the other physical properties of non-Schwarzschild-like black hole solutions of the theory is also crucial.

\section*{Acknowledgements}
This research has received funding support from the NSRF via the Program Management Unit for Human Resource and Institutional Development, Research and Innovation grant number $B13F670063$.

\appendix
\section{Explicit Terms in the Continued Fraction Approximation}\label{sec:Appendix}

We present terms up to {fourth order} in the continued fraction approximation \eqref{cfrac}: 
\begin{align}
\epsilon&=-\dfrac{F_1}{r_+}-1,\,\,\,\, a_1=-1-a_{0}+2\epsilon+r_{+}h_1,\,\,\,\, a_{2}=-\dfrac{1}{ a_1} \left[4a_1-5\epsilon+1+3 a_{0}+ h_{2}{r_+}^2\right]
\nonumber \\
a_{4}&=-\dfrac{1}{a_1a_{2}a_{3}}\Bigg[h_{4}r_{+}^2+a_1a_{2}^3+2a_1a_{3}a_{2}^2+a_1a_{2}a_{3}^2+6a_1a_{2}^2+6a_1a_{2}a_{3}+15a_1a_{2}+10a_{0}\nonumber\\
    &~~~+20a_1-14\epsilon +1\Bigg],\;\;\;
a_{3}=-\dfrac{1}{{a_1}{a_{2}}}\Big[-{h_{3}}{r_+}^3+{a_1}{{a_{2}}}^{2}+5{a_1}{a_{2}}+6{a_{0}}+10{a_1}-9\epsilon+1\Big],
\end{align}
and 
\begin{align} %\label{bb34}
b_1 &= -1+\sqrt{\dfrac{h_1}{f_1}},\,\,\,\,\,\,\,\,\,  b_{2}=\dfrac{(-4f_{1}+f_{2}r_{+})b_{1}^2+2(-2f_{1}+f_{2}r_{+})b_{1}+r_{+}(f_{2}-h_{2})}{2f_{1}b_{1}(1+b_{1})},\,\,\nonumber\\
b_{3} &= \dfrac{1}{2f_{1}b_{1}b_{2}(1+b_{1})}\Big[(-f_{3}r_{+}^2+2f_{2}r_{+}(2+b_{2})-f_{1}(10+3b_{2}^2+10b_{2}))b_{1}^2+(-2f_{3}r_{+}^2+2f_{2}r_{+}(2+b_{2}\nonumber\\
&-2f_{1}(3+b_{2}^2+3b_{2})))b_{1}-r_{+}^2(f_{3}-h_{3})\Big]\nonumber\\
b_{4} &=\dfrac{1}{2f_{1}b_{1}b_{2}b_{3}}\Big[(-4f_{1}b_{2}^3+b_{2}^{2}(3f_{2}r_{+}-6f_{1}(b_{3}+3))+b_{2}(-2f_{3}r_{+}^{2}+2f_{2}r_{+}(b_{3}+5)-2f_{1}(6b_{3}+b_{3}^2+15))b_{2}\nonumber\\
&+f_{4}r_{+}^3-20f_{1}-4f_{3}r_{+}^2+10f_{2}r_{+})b_{1}^2+(-2f_{1}b_{2}^3+(2f_{2}r_{+}-4f_{1}(b_{3}+2))b_{2}^2+(-2f_{3}r_{+}^2+2f_{2}r_{+}(b_{3}+3)\nonumber\\
&-2f_{1}(6+4b_{3}+b_{3}^2))b_{2}-8f_{1}+6f_{2}r_{+}-4f_{3}r_{+}^2+2f_{4}r_{+}^3)b_{1}+r_{+}^{3}(f_{4}-h_{4})\Big].  \nonumber
\end{align}
The quantities  $f_3$ and $h_3$ are respectively given in \eqref{eq9}
and
\begin{align}
f_{3} &= \dfrac{1}{288\alpha^{2}r_{+}^3 f_{1}^2}\Bigg[544\alpha^2 r_{+}f_{1}^{4}-4\alpha(64\alpha+5r_{+}^3)f_{1}^{3}+r_{+}(r_{+}^2-28\alpha)f_{1}^2+8(6\alpha +r_{+}^2)f_{1}-9r_{+}\Bigg],\nonumber \\
h_{3} &= \dfrac{h_{1}}{288\alpha^2 r_{+}^3f_{1}^4}\Bigg[9r_{+}+48\alpha f_{1}-52\alpha r_{+} f_{1}^2+4\alpha r_{+}^2f_{1}^3-640\alpha^2 f_{1}^{3}+928\alpha^2 r_{+}f_{1}^{4}-16r_{+}^2f_{1}+7r_{+}^3f_{1}^2\Bigg],\nonumber\\
f_{4}&=\dfrac{1}{4608r_{+}^{5}\alpha^3 f_{1}^{5}}\Bigg[-10752\alpha^3 f_{1}^6 r_{+}^3+32r_{+}\alpha^2f_{1}^{5}(35r_{+}^2+232\alpha)+2\alpha f_{1}^4(17r_{+}^4-640\alpha^2-880\alpha r_{+}^2)\nonumber\\
&-2r_{+}(r_{+}^4-74\alpha r_{+}^2-560\alpha^2)f_{1}^3-2f_{1}^2(8r_{+}^4+240\alpha^2+199\alpha r_{+}^2)+9r_{+}f_{1}(24\alpha+5r_{+}^2)-27r_{+}^2\Bigg],\nonumber\\
h_{4}&=-\dfrac{h_{1}}{4608\alpha^3 r_{+}^5f_{1}^6}\Bigg[-45r_{+}^2+928\alpha^2f_{1}^5r_{+}^3+332\alpha r_{+}^{3}f_{1}^3+18944\alpha^3 f_{1}^6 r_{+}^2-2464\alpha^2 r_{+}^2f_{1}^4-\nonumber\\
&15616\alpha^3r_{+}f_{1}^{5}+1280\alpha^3 f_{1}^4+1056\alpha^2r_{+}f_{1}^3+480\alpha^2 f_{1}^2-120\alpha f_{1}r_{+}-194\alpha r_{+}^4f_{1}^4-18\alpha r_{+}^2f_{1}^2\nonumber\\
&+115f_{1}r_{+}^3-96r_{+}^4f_{1}^2+26r_{+}^5f_{1}^3\Bigg].
\end{align}

\section{Coefficients of continued fraction expansion of the \\ non-Schwarzschild background}\label{appE}
In this Appendix, based on the analytical work on quadratic gravity in the paper \cite{Kokkotas:2017zwt}, the coefficients of continued fraction expansion on the non-GR solution are given as
\begin{align}\label{epsilon1}
    \epsilon \approx & (1054-1203p)\left(\dfrac{3}{1271}+\dfrac{p}{1529}\right),\;\;\;\;
    a_{1}\approx  (1054-1203p)\left(\dfrac{7}{1746}-\dfrac{5p}{2421}\right), \\
    a_{2}\approx & \dfrac{6p^2}{17}+\dfrac{5p}{6}-\dfrac{131}{102},\;\;\;\;
    a_{3}\approx \dfrac{-385p+\dfrac{9921p^2}{31}+\dfrac{4857}{29}}{237-223p},\;\;\;
    a_{4}\approx \dfrac{\dfrac{9p^2}{14}+\dfrac{3149p}{42}-\dfrac{2803}{14}}{237-223p},  
\end{align}
and
\begin{align}
b_{1}&=(1054-1203p)\left(\dfrac{p}{1465}-\dfrac{2}{1585}\right),\;\;\;b_{2}=\dfrac{81p^2}{242}-\dfrac{109p}{118}-\dfrac{16}{89},\\
b_{3}&=-\dfrac{2p}{57}+\dfrac{29}{56},\;\;\;\;\;b_{4}=\dfrac{13p}{95}-\dfrac{121}{98},
\end{align}
here $p=r_{+}/\sqrt{2\alpha}$.
From the definition of $\epsilon$ from equation \eqref{epsilon1}, the mass is obtained as follows
\begin{eqnarray}
    M=\dfrac{r_+}{2}\left(1+\epsilon\right)
\end{eqnarray}
It should be remarked that to plot the mass in terms of $r_{+}$ in figure \ref{mrplotta1}, we have used the above definition for the mass, i.e,
\begin{eqnarray}
    M=\dfrac{r_+}{2}\left[1+(1054-1203p)\left(\dfrac{3}{1271}+\dfrac{p}{1529}\right)\right].
\end{eqnarray}

\bibliographystyle{unsrt}
\bibliography{refs}

\providecommand{\noopsort}[1]{}\providecommand{\singleletter}[1]{#1}%
\begin{thebibliography}{10}

\bibitem{PhysRevD.16.953}
K.~S. Stelle.
\newblock Renormalization of higher-derivative quantum gravity.
\newblock {\em Phys. Rev. D}, 16:953--969, Aug 1977.

\bibitem{Lu:2015cqa}
H.~Lu, A.~Perkins, C.~N. Pope, and K.~S. Stelle.
\newblock {Black Holes in Higher-Derivative Gravity}.
\newblock {\em Phys. Rev. Lett.}, 114(17):171601, 2015.

\bibitem{PhysRevD.92.124019}
H.~L\"u, A.~Perkins, C.~N. Pope, and K.~S. Stelle.
\newblock Spherically symmetric solutions in higher-derivative gravity.
\newblock {\em Phys. Rev. D}, 92:124019, Dec 2015.

\bibitem{Rezzolla:2014mua}
Luciano Rezzolla and Alexander Zhidenko.
\newblock {New parametrization for spherically symmetric black holes in metric theories of gravity}.
\newblock {\em Phys. Rev. D}, 90(8):084009, 2014.

\bibitem{Podolsky:2018pfe}
Jiri Podolsky, Robert Svarc, Vojtech Pravda, and Alena Pravdova.
\newblock {Explicit black hole solutions in higher-derivative gravity}.
\newblock {\em Phys. Rev. D}, 98(2):021502, 2018.

\bibitem{Bonanno:2019rsq}
Alfio Bonanno and Samuele Silveravalle.
\newblock {Characterizing black hole metrics in quadratic gravity}.
\newblock {\em Phys. Rev. D}, 99(10):101501, 2019.

\bibitem{Sajadi:2020axg}
S.~N. Sajadi, Robert~B. Mann, N.~Riazi, and Saeed Fakhry.
\newblock {Analytically Approximation Solution to Higher Derivative Gravity}.
\newblock 10 2020.

\bibitem{Sajadi:2022ybs}
S.~N. Sajadi and S.~H. Hendi.
\newblock {Analytically approximation solution to Einstein-Cubic gravity}.
\newblock {\em Eur. Phys. J. C}, 82(8):675, 2022.

\bibitem{Sajadi:2022tgi}
Seyed~Naseh Sajadi, Ali Hajilou, and Seyed~Hossein Hendi.
\newblock {Analytically approximation solution to $R^{2}$ gravity}.
\newblock {\em Eur. Phys. J. C}, 83(1):45, 2023.

\bibitem{Kokkotas:2017zwt}
K.~Kokkotas, R.~A. Konoplya, and A.~Zhidenko.
\newblock {Non-Schwarzschild black-hole metric in four dimensional higher derivative gravity: analytical approximation}.
\newblock {\em Phys. Rev. D}, 96:064007, 2017.

\bibitem{Kokkotas:2017ymc}
K.~D. Kokkotas, R.~A. Konoplya, and A.~Zhidenko.
\newblock {Analytical approximation for the Einstein-dilaton-Gauss-Bonnet black hole metric}.
\newblock {\em Phys. Rev. D}, 96(6):064004, 2017.

\bibitem{Konoplya:2022iyn}
R.~A. Konoplya.
\newblock {Quasinormal modes in higher-derivative gravity: Testing the black hole parametrization and sensitivity of overtones}.
\newblock {\em Phys. Rev. D}, 107(6):064039, 2023.

\bibitem{Antoniou:2024jku}
Georgios Antoniou, Leonardo Gualtieri, and Paolo Pani.
\newblock {Gravitational quasinormal modes in quadratic gravity}.
\newblock 12 2024.

\bibitem{Konoplya:2011qq}
R.~A. Konoplya and A.~Zhidenko.
\newblock {Quasinormal modes of black holes: From astrophysics to string theory}.
\newblock {\em Rev. Mod. Phys.}, 83:793--836, 2011.

\bibitem{Kokkotas:1999bd}
Kostas~D. Kokkotas and Bernd~G. Schmidt.
\newblock {Quasinormal modes of stars and black holes}.
\newblock {\em Living Rev. Rel.}, 2:2, 1999.

\bibitem{Promsiri:2023yda}
Chatchai Promsiri, Takol Tangphati, Ekapong Hirunsirisawat, and Supakchai Ponglertsakul.
\newblock {Scalarization of planar anti\textendash{}de Sitter charged black holes in Einstein-Maxwell-scalar theory}.
\newblock {\em Phys. Rev. D}, 108(2):024015, 2023.

\bibitem{Leaver:1985ax}
E.~W. Leaver.
\newblock {An Analytic Representation for the Quasi-Normal Modes of Kerr Black Holes}.
\newblock {\em Procedings of the Royal Society of London. Series A, Mathematical and Physical Sciences}, 402(1823):285--298, 1985.

\bibitem{Cho:2009cj}
H.~T. Cho, A.~S. Cornell, Jason Doukas, and Wade Naylor.
\newblock {Black hole quasinormal modes using the asymptotic iteration method}.
\newblock {\em Class. Quant. Grav.}, 27:155004, 2010.

\bibitem{Burikham:2017gdm}
Piyabut Burikham, Supakchai Ponglertsakul, and Lunchakorn Tannukij.
\newblock {Charged scalar perturbations on charged black holes in de Rham-Gabadadze-Tolley massive gravity}.
\newblock {\em Phys. Rev. D}, 96(12):124001, 2017.

\bibitem{Ponglertsakul:2018smo}
Supakchai Ponglertsakul, Piyabut Burikham, and Lunchakorn Tannukij.
\newblock {Quasinormal modes of black strings in de Rham\textendash{}Gabadadze\textendash{}Tolley massive gravity}.
\newblock {\em Eur. Phys. J. C}, 78(7):584, 2018.

\bibitem{Ponglertsakul:2020ufm}
Supakchai Ponglertsakul and Bogeun Gwak.
\newblock {Massive scalar perturbations on Myers-Perry\textendash{}de Sitter black holes with a single rotation}.
\newblock {\em Eur. Phys. J. C}, 80(11):1023, 2020.

\bibitem{Jansen:2017oag}
Aron Jansen.
\newblock {Overdamped modes in Schwarzschild-de Sitter and a Mathematica package for the numerical computation of quasinormal modes}.
\newblock {\em Eur. Phys. J. Plus}, 132(12):546, 2017.

\bibitem{Schutz:1985km}
Bernard~F. Schutz and Clifford~M. Will.
\newblock {BLACK HOLE NORMAL MODES: A SEMIANALYTIC APPROACH}.
\newblock {\em Astrophys. J. Lett.}, 291:L33--L36, 1985.

\bibitem{Iyer:1986np}
Sai Iyer and Clifford~M. Will.
\newblock {Black Hole Normal Modes: A {WKB} Approach. 1. Foundations and Application of a Higher Order {WKB} Analysis of Potential Barrier Scattering}.
\newblock {\em Phys. Rev. D}, 35:3621, 1987.

\bibitem{Konoplya:2003ii}
R.~A. Konoplya.
\newblock {Quasinormal behavior of the d-dimensional Schwarzschild black hole and higher order WKB approach}.
\newblock {\em Phys. Rev. D}, 68:024018, 2003.

\bibitem{Matyjasek:2017psv}
Jerzy Matyjasek and Micha\l{} Opala.
\newblock {Quasinormal modes of black holes. The improved semianalytic approach}.
\newblock {\em Phys. Rev. D}, 96(2):024011, 2017.

\bibitem{Matyjasek:2019eeu}
Jerzy Matyjasek and Malgorzata Telecka.
\newblock {Quasinormal modes of black holes. II. Pad\'e summation of the higher-order WKB terms}.
\newblock {\em Phys. Rev. D}, 100(12):124006, 2019.

\bibitem{Konoplya:2019hlu}
R.~A. Konoplya, A.~Zhidenko, and A.~F. Zinhailo.
\newblock {Higher order WKB formula for quasinormal modes and grey-body factors: recipes for quick and accurate calculations}.
\newblock {\em Class. Quant. Grav.}, 36:155002, 2019.

\bibitem{Tangphati:2023xnw}
Takol Tangphati, Menglong Youk, and Supakchai Ponglertsakul.
\newblock {Magnetically charged regular black holes in $f(R,T)$ gravity coupled to nonlinear electrodynamics}.
\newblock 12 2023.

\bibitem{Gogoi:2024vcx}
Dhruba~Jyoti Gogoi and Supakchai Ponglertsakul.
\newblock {Constraints on Quasinormal modes from Black Hole Shadows in regular non-minimal Einstein Yang-Mills Gravity}.
\newblock 2 2024.

\bibitem{Bhar:2024ehw}
Piyali Bhar, Dhruba~Jyoti Gogoi, and Supakchai Ponglertsakul.
\newblock {Noncommutative black hole in de Rham-Gabadadze-Tolley like massive gravity}.
\newblock 4 2024.

\bibitem{vf}
Valeria Ferrari and Bahram Mashhoon.
\newblock New approach to the quasinormal modes of a black hole.
\newblock {\em Phys. Rev. D}, 30:295--304, Jul 1984.

\bibitem{Cardoso:2003sw}
Vitor Cardoso and Jose P.~S. Lemos.
\newblock {Quasinormal modes of the near extremal Schwarzschild-de Sitter black hole}.
\newblock {\em Phys. Rev. D}, 67:084020, 2003.

\bibitem{Wuthicharn:2019olp}
Taum Wuthicharn, Supakchai Ponglertsakul, and Piyabut Burikham.
\newblock {Quasi-normal modes of near-extremal black holes and black strings in massive gravity background}.
\newblock {\em Int. J. Mod. Phys. D}, 31(01):2150127, 2022.

\bibitem{Burikham:2020dfi}
Piyabut Burikham, Supakchai Ponglertsakul, and Taum Wuthicharn.
\newblock {Quasi-normal modes of near-extremal black holes in generalized spherically symmetric spacetime and strong cosmic censorship conjecture}.
\newblock {\em Eur. Phys. J. C}, 80(10):954, 2020.

\bibitem{Churilova:2021nnc}
M.~S. Churilova, R.~A. Konoplya, and A.~Zhidenko.
\newblock {Analytic formula for quasinormal modes in the near-extreme Kerr-Newman\textendash{}de Sitter spacetime governed by a non-P\"oschl-Teller potential}.
\newblock {\em Phys. Rev. D}, 105(8):084003, 2022.

\bibitem{Wald1}
Robert~M. Wald.
\newblock {Black hole entropy is the Noether charge}.
\newblock {\em Phys. Rev. D}, 48(8):R3427--R3431, 1993.

\bibitem{Wald2}
Vivek Iyer and Robert~M. Wald.
\newblock {Some properties of Noether charge and a proposal for dynamical black hole entropy}.
\newblock {\em Phys. Rev. D}, 50:846--864, 1994.

\bibitem{Hajian:2023bhq}
Kamal Hajian and Bayram Tekin.
\newblock {Coupling Constants as Conserved Charges in Black Hole Thermodynamics}.
\newblock {\em Phys. Rev. Lett.}, 132(19):191401, 2024.

\bibitem{Lu:2017kzi}
Hong L\"u, A.~Perkins, C.~N. Pope, and K.~S. Stelle.
\newblock {Lichnerowicz Modes and Black Hole Families in Ricci Quadratic Gravity}.
\newblock {\em Phys. Rev. D}, 96(4):046006, 2017.

\bibitem{Mashhoon:1985cya}
Bahram Mashhoon.
\newblock Stability of charged rotating black holes in the eikonal approximation.
\newblock {\em Phys. Rev. D}, 31:290--293, Jan 1985.

\bibitem{Mafuz:2023qxn}
Md.~Golam Mafuz, Rishank Diwan, Soumya Jana, and Sayan Kar.
\newblock {Shadows of a generic class of spherically symmetric, static spacetimes}.
\newblock {\em Eur. Phys. J. Plus}, 139(3):219, 2024.

\bibitem{Jusufi:2019ltj}
Kimet Jusufi.
\newblock {Quasinormal Modes of Black Holes Surrounded by Dark Matter and Their Connection with the Shadow Radius}.
\newblock {\em Phys. Rev. D}, 101(8):084055, 2020.

\bibitem{Jusufi:2020dhz}
Kimet Jusufi.
\newblock {Connection Between the Shadow Radius and Quasinormal Modes in Rotating Spacetimes}.
\newblock {\em Phys. Rev. D}, 101(12):124063, 2020.

\bibitem{Hendi:2020knv}
S.~H. Hendi, S.~N. Sajadi, and M.~Khademi.
\newblock {Physical properties of a regular rotating black hole: Thermodynamics, stability, and quasinormal modes}.
\newblock {\em Phys. Rev. D}, 103(6):064016, 2021.

\bibitem{Konoplya:2022kld}
R.~A. Konoplya and A.~Zhidenko.
\newblock {How general is the strong cosmic censorship bound for quasinormal modes?}
\newblock {\em JCAP}, 11:028, 2022.

\end{thebibliography}
\end{document}